\documentstyle[preprint,aps,epsf,floats]{revtex}
\begin{document}

\tightenlines

\def\lqcd{\Lambda_{\rm QCD}}
\def\xslash#1{{\rlap{$#1$}/}}
\def\dsl{\,\raise.15ex\hbox{/}\mkern-13.5mu D}
\preprint{\vbox{\hbox{UTPT--99-19} 
\hbox{hep-ph/9911404}}}

\def\ctp#1#2#3{\CTP{\bf #1} (#2) #3}
\def\jetpl#1#2#3{\JETPL{\bf #1} (#2) #3}
\def\nc#1#2#3{\NC{\bf #1} (#2) #3}
\def\np#1#2#3{\NP{\bf B#1} (#2) #3}
\def\pl#1#2#3{\PL B {\bf #1} (#2) #3}
\def\prl#1#2#3{\PRL{\bf #1} (#2) #3}
\def\prd#1#2#3{\PR D {\bf #1} (#2) #3}
\def\prep#1#2#3{\PRep{\bf #1} (#2) #3}
\def\physrev#1#2#3{\PR{\bf #1} (#2) #3}
\def\sjnp#1#2#3{\SJNP{\bf #1} (#2) #3}
\def\nuvc#1#2#3{\NC{\bf #1A} (#2) #3}
\def\blankref#1#2#3{   {\bf #1} (#2) #3}
\def\ibid#1#2#3{{\it ibid,\/}  {\bf #1} (#2) #3}
\def\AP{{\it Ann.\ Phys.\ }}
\def\CMP{{\it Comm.\ Math.\ Phys.\ }}
\def\CTP{{\it Comm.\ Theor.\ Phys.\ }}
\def\IJMP{{\it Int.\ Jour.\ Mod.\ Phys.\ }}
\def\JETPL{{JETP Lett.\ }}
\def\NC{{\it Nuovo Cimento\ }}
\def\NP{{Nucl.\ Phys.\ }}
\def\PL{{Phys.\ Lett.\ }}
\def\PR{{Phys.\ Rev.\ }}
\def\PRep{{Phys.\ Rep.\ }}
\def\PRL{{Phys.\ Rev.\ Lett.\ }}

\title{Nonperturbative corrections to moments of the decay $B \to X_s
  \ell^+ \ell^-$} 

\author{Christian~W.~Bauer\footnote{{\tt bauer@physics.utoronto.ca}}, 
        Craig~N.~Burrell\footnote{{\tt burrell@physics.utoronto.ca}}}

\address{
\medskip Department of Physics, University of Toronto\\60
  St.~George Street, Toronto, Ontario,
  Canada M5S 1A7 
\medskip}

\bigskip
\date{November 1999} 

\maketitle

\begin{abstract} We study nonperturbative corrections to the inclusive rare 
  decay $B \to X_s \ell^+ \ell^-$ by performing an
  operator product expansion (OPE) to ${\cal O}(1/m_b^3)$. 
  The values of the matrix elements 
  entering at this order are unknown and introduce uncertainties into
  physical quantities.
  We study uncertainties introduced into the partially integrated rate,
  moments of the hadronic spectrum, as well as the forward-backward
  asymmetry.  We find that for large dilepton invariant mass $q^2 >
  M_{\psi'}^2$ these uncertainties are large.  We also assess the
  possibility of  
  extracting the HQET parameters $\lambda_1$ and $\bar{\Lambda}$ using 
  data from this process.
\end{abstract}

\pagebreak

\section{Introduction}
Rare $B$ decays mediated via flavour changing neutral currents have
received much attention because of their sensitivity to physics
beyond the standard model. In the standard model these decays occur
via penguin and box diagrams with virtual electroweak bosons and up-type 
quarks in the loops. Because of the large top quark mass, 
the contribution with a top quark in the loop dominates. At energy scales 
below the mass of the top quark and $W$ boson, 
it is convenient to switch to an effective theory where the top quark and
the weak bosons have been integrated out of the theory. The $b \to s$
transition is then mediated by the effective Hamiltonian 
\cite{grinstein_savage_wise}
\begin{equation}\label{eff_hamilton}
{\cal H}_{eff} = -\frac{4 G_F}{\sqrt{2}} V_{tb}V_{ts}^* \sum_{i=1}^{10} 
                  C_i(\mu) {\cal O}_i(\mu),
\end{equation}
where the operators are commonly defined by
\begin{eqnarray}
  \label{eq:operatorbasis}
  O_1 &=& (\bar{s}_{L \alpha} \gamma_\mu b_{L \alpha})
  (\bar{c}_{L \beta} \gamma^\mu c_{L \beta}), \nonumber   \\
  O_2 &=& (\bar{s}_{L \alpha} \gamma_\mu b_{L \beta})
  (\bar{c}_{L \beta} \gamma^\mu c_{L \alpha}), \nonumber   \\
  O_3 &=& (\bar{s}_{L \alpha} \gamma_\mu b_{L \alpha})
  \sum_{q=u,d,s,c,b}
  (\bar{q}_{L \beta} \gamma^\mu q_{L \beta}), \nonumber   \\
  O_4 &=& (\bar{s}_{L \alpha} \gamma_\mu b_{L \beta})
  \sum_{q=u,d,s,c,b}
  (\bar{q}_{L \beta} \gamma^\mu q_{L \alpha}), \nonumber   \\
  O_5 &=& (\bar{s}_{L \alpha} \gamma_\mu b_{L \alpha})
  \sum_{q=u,d,s,c,b}
  (\bar{q}_{R \beta} \gamma^\mu q_{R \beta}), \nonumber   \\
  O_6 &=& (\bar{s}_{L \alpha} \gamma_\mu b_{L \beta})
  \sum_{q=u,d,s,c,b}
  (\bar{q}_{R \beta} \gamma^\mu q_{R \alpha}), \nonumber   \\  
  O_7 &=& \frac{e}{16 \pi^2}
  \bar{s}_{\alpha} \sigma_{\mu \nu} (m_b R + m_s L) b_{\alpha}
  F^{\mu \nu},    \nonumber                               \\
  O_8 &=& \frac{g}{16 \pi^2}
  \bar{s}_{\alpha} T_{\alpha \beta}^a \sigma_{\mu \nu} (m_b R + m_s L)  
  b_{\beta} G^{a \mu \nu},   \nonumber\\
  O_9 &=& \frac{e^2}{16 \pi^2} \bar{s}_\alpha \gamma^{\mu} L b_\alpha
  \bar{\ell} \gamma_{\mu} \ell , \nonumber\\
  O_{10} &=& \frac{e^2}{16 \pi^2} \bar{s}_\alpha \gamma^{\mu} L
  b_\alpha \bar{\ell} \gamma_{\mu}\gamma_5 \ell.
\end{eqnarray}
Here $L/R = \frac{1}{2} \left( 1 \pm \gamma^5 \right)$ are the usual
left and right handed chiral projection operators. The values of the
Wilson coefficients $C_i(m_b)$ have been calculated in 
the next to
leading log approximation \cite{misiak,buras-munz} in the standard
model and are given in
Table \ref{Wilson_coefficients}. 

\begin{table}[htbp]
\label{Wilson_coefficients}
\begin{tabular}{|ccccccccc|}
$C_1$ & $C_2$ & $C_3$ & $C_4$ & $C_5$ & $C_6$ & $C_7$ & $C_9$ & $C_{10}$\\
-0.240 & 1.103 & 0.011 & -0.025 & 0.007 & -0.030 & -0.311 & 4.153 & -4.546
\end{tabular}
\caption{The Wilson coefficients $C_i(m_b)$ in the next-to-leading
log approximation.} 
\end{table}

Physics beyond the standard model will generally introduce new
contributions to the loop and will therefore
modify the values of these coefficients
\cite{newphysics}. Measurements of the Wilson coefficients may 
therefore indirectly constrain new physics scenarios.  For example,
the decay $b \to s
\gamma$ is proportional to $|C_7|^2$, and the recent measurements of the 
branching ratio $B
\to K^* \gamma$ \cite{CLEOex} and inclusive rate $B \to X_s \gamma$ 
\cite{CLEOin} have placed constraints on models of physics beyond the 
standard model which modify the magnitude of $C_7$ \cite{constraints}. 

The 
decay $b \to s \ell^+ \ell^-$ is suppressed, relative to $b \to s \gamma$, 
by an additional factor of
the electromagnetic coupling constant and has not yet been
observed \cite{experimental_search}. It has, however, the appeal of 
being sensitive to the signs and magnitudes of 
$C_7$, $C_9$, and $C_{10}$, making it a potentially
more powerful probe than $b \to s \gamma$ of beyond the standard 
model physics. Experimental studies of this process impose cuts on the
available phase space.
This is primarily due to the necessity of removing the resonance
from $B \to (J/\psi,\psi') X_s$ 
with the $(J/\psi,\psi')$ decaying into two leptons.  We incorporate 
representative cuts into the theoretical analysis.
  
Using an operator product expansion (OPE) several observables of the
inclusive decay $B \to X_s \ell^+ \ell^-$ 
have been calculated including the leading non-perturbative
corrections \cite{ahhm_btosee_lepton,ah_btosee_hadron}.  
In this framework these leading   
corrections arise as matrix elements of dimension five operators, suppressed by
two powers of the $b$ quark mass, and are conventionally parameterized
by two quantities, $\lambda_1$ and 
$\lambda_2$.  A third parameter $\bar{\Lambda}$  
enters through the difference of the $b$ quark mass and $B$ meson mass
\begin{equation} 
\label{mass_relation_2} 
m_b = M_B - \bar{\Lambda} + \frac{\lambda_1 + 3 \lambda_2}{2 m_b} + \cdot\cdot\cdot \nonumber.  
\end{equation}
Whereas $\lambda_2$ can be determined from the $B^*-B$
mass splitting, $\lambda_2 =  
(M_{B^*}^2 - M_B^2)/4 \simeq 0.12 \; \mbox{GeV}^2$, no such simple relation
exists for $\lambda_1$.  It has been 
estimated using various methods to lie in the range $0.1 \; \mbox{GeV}^2 \le
(-\lambda_1) \le 0.6 \; \mbox{GeV}^2$ \cite{lam1values}.  
  
In a previous paper we extended the analysis of the total rate for $B
\to X_s \ell^+ \ell^-$ to
one order higher in the OPE \cite{bauerburrell}.    
The dimension six operators arising at this order can be parametrized
by six quantities, commonly labelled 
$\rho_{1-2}$ and ${\cal T}_{1-4}$, all of which are unknown.  We found
that the uncertainties introduced  
by these six parameters can be significant, depending primarily on the 
actual values of the matrix elements and the amount of accessible phase
space.  In this paper we give the details of that analysis and also 
present calculations for the forward-backward asymmetry and moments 
of the hadron invariant mass spectrum at ${\cal O}(1/m_b^3)$. 
As in our 
previous analysis, we neglect perturbative
effects and effects due to the finite mass of the $s$ quark,
which have been considered elsewhere \cite{ah_btosee_hadron}. 

It has also been proposed that, rather than use this decay to search for
new physics, it might instead be used to  
extract the parameters $\bar{\Lambda}$ and $\lambda_1$ through a 
measurement of its hadronic invariant mass moments \cite{ah_btosee_hadron}.  
We estimate the uncertainties in this
extraction due to the unknown matrix elements of the dimension six operators. 

This paper is organized as follows. In
section \ref{section_formalism} we briefly introduce the formalism used
to calculate the nonperturbative corrections, and we present the results
for the decay rate in section \ref{section_partial}. In section 
\ref{section_fbasym} we calculate the forward-backward asymmetry of
the lepton pair.  We then
proceed in section \ref{section_moments} to calculate moments of the
hadronic invariant mass spectrum and estimate uncertainties in
extracting $\bar{\Lambda}$ and $\lambda_1$ from these moments. 
Finally we discuss the results and state our conclusions.

\section{Operator Product Expansion and Kinematics}
\label{section_formalism}
The procedure for calculating nonperturbative contributions to
heavy hadron decays has been thoroughly discussed in the
literature \cite{chay,inclusive}, and we present here only a brief outline of the
technique. The differential rate is proportional to the product of a 
lepton tensor
$L_{\mu\nu}$ and a hadron tensor $W^{\mu\nu}$ and for the process
in question it may be written as
\begin{equation}
d\Gamma = \frac{1}{2M_B} \frac{G_F^2 \alpha^2}{2\pi^2} |V_{ts}^* V_{tb}|^2 
d\Pi \left( L_{\mu\nu}^L W^{L\mu\nu} + L_{\mu\nu}^R W^{R\mu\nu} \right)
\end{equation}
where $\Pi$ denotes the three body phase space.  The spin-summed 
tensor $L_{\mu\nu}$ for massless leptons is
\begin{equation}
L_{L(R)}^{\mu\nu} = 2 \left[ p_+^\mu p_-^\nu + p_-^\mu  p_+^\nu  - 
g^{\mu\nu} p_+\cdot p_- \mp i \epsilon^{\mu\nu\alpha\beta}p_{+\alpha}
p_{-\beta} \right].
\end{equation}
The hadron tensor $W^{\mu\nu}$ is related via the optical theorem to the 
imaginary part of the forward scattering matrix element 
$W^{\mu\nu} = (-1/\pi)\,{\rm Im} T^{\mu\nu}$ where  
\begin{equation} 
\label{discontinuity}
T^{L(R)}_{\mu\nu} = - i \int d^4 x \, e^{-i q \cdot x} 
  \left\langle B \left| T\{ J^{L(R)^\dagger}_\mu(x) ,J^{L(R)}_\nu(0)\}
  \right| B \right\rangle.
\end{equation} 
In this equation $J^\mu$ denotes the current mediating this
transition, and is given by
\begin{equation}
J^\mu_{L(R)} = \bar{s} \left[ R \gamma^\mu \left( C_9^{eff} \mp C_{10} + 
  2 C_7^{eff} \frac{ \rlap /\hat{q} }{\hat{q}^2} \right) + 
  2 \hat{m}_s C_7^{eff} \gamma^\mu \frac{ \rlap /\hat{q} }{\hat{q}^2} L
  \right] b
\end{equation}
where $q \equiv (p_+ + p_-)$ is the dilepton
momentum\footnote{
Notice that this
current $J_\mu$ reduces to the $(V-A)$ current when 
$C_7^{eff} = 0$, $C_9^{eff} = 1/2$, and $C_{10} = -1/2$. 
This provides some useful cross-checks with 
known results for semileptonic $B$ decays \cite{gremm}.}.
In accordance with convention, we have defined two effective 
Wilson coefficients: 
$C_7^{eff} \equiv C_7 - C_5/3 - C_6$ and $C_9^{eff}$. The latter contains 
the operator mixing of $O_{1-6}$ into $O_9$ as well as the one loop
matrix elements of $O_{1-6,9}$ \cite{misiak,buras-munz}.
The full analytic expression for $C_9^{eff}$ is quite
lengthy and may be found in \cite{buras-munz}. 

Since in the decay of a $b$ quark the momentum transfer to the final state parton
is large, the time--ordered product (\ref{discontinuity}) 
can be expanded in terms of local operators \cite{chay,inclusive}
\begin{equation}
\label{schematicOPE}
- i \int d^4 x \, e^{-i q \cdot x} 
  T\{ J^\dagger(x) ,J(0)\} \sim \frac{1}{m_b} \left[ {\cal O}_0 +
    \frac{1}{2m_b} {\cal O}_1 +
  \frac{1}{4m_b^2} {\cal O}_2 + \frac{1}{8m_b^3} {\cal O}_3 + \ldots
\right], 
\label{ope}
\end{equation}
where ${\cal O}_n$ represents a set of local operators of dimension
$d=(3+n)$, each operator containing $n$ derivatives. 
For a generic current $J_\mu$ the expressions for these
operators are quite lengthy. The complete set of operators for $d \le 5$ 
\cite{fls1994} and $d=6$ \cite{christian} appear in the literature.  In
this study we include operators up to and including dimension $d=6$. 

The standard Lorentz decomposition for the forward scattering amplitude is
\begin{equation}\label{eq:hadrontensor}
  T_{\mu \nu} = -T_1 \, g_{\mu \nu} + T_2 \, v_\mu \, v_\nu 
  + T_3 \, i \epsilon_{\mu \nu \alpha \beta} \, 
  v^\alpha \, \hat{q}^\beta + T_4 \, \hat{q}^\mu \, \hat{q}^\nu + 
  T_5 (\hat{q}^\mu \, v^\nu + \hat{q}^\nu v^\mu),
\end{equation}
where $v^\mu$ is the four--velocity of the initial $b$ quark $p_b = m_b v$.
Since in this paper we treat the final state leptons as massless $(\ell = e,\mu)$,
the form factors $T_{4-5}$ do not contribute to observables.

It is clear from (\ref{discontinuity}) and (\ref{schematicOPE}) that to 
calculate these form factors we must take matrix
elements of the operators ${\cal O}_n$. Matrix elements 
of dimension four operators
vanish at leading order in the $1/m_b$ expansion \cite{chay} and
matrix elements of dimension five operators may be
parameterized by $\lambda_1$ and $\lambda_2$ \cite{paramd2} 
\begin{equation}
  \langle B(v) | \bar{h}_v \Gamma iD_\mu iD_\nu h_v | B(v) \rangle = M_B {\rm
  Tr} \left\{\Gamma P_+ \left(\frac{1}{3} \lambda_1 (g_{\mu\nu} - v_\mu v_\nu)
  +\frac{1}{2} \lambda_2 i \sigma_{\mu\nu}\right) P_+\right\},
\label{lambda_gen}
\end{equation}
where $P_+=\frac{1}{2}(1+\rlap/v)$, and $\Gamma$ is an arbitrary Dirac 
structure. 

Finally, the dimension six operators may
be parameterized by the matrix elements of two local operators 
\cite{gremm,mannel} 
\begin{eqnarray}\label{rho}
\frac{1}{2M_B}\langle B(v)| \bar{h}_v iD_\alpha iD_\mu iD_\beta
h_v | B(v)\rangle&=&\frac{1}{3}\rho_1\left(g_{\alpha\beta}-v_\alpha
v_\beta\right) v_\mu, \nonumber\\
 \frac{1}{2M_B}\langle B(v)| \bar{h}_v
iD_\alpha iD_\mu iD_\beta \gamma_\delta \gamma_5 h_v
| B(v)\rangle&=&\frac{1}{2} \rho_2 \, i\epsilon_{\nu\alpha\beta\delta}
v^\nu v_\mu 
\end{eqnarray}
and by matrix elements of two time--ordered products
\begin{eqnarray}\label{tau}
\frac{1}{2 M_B} \langle B(v)|\bar{h}_v (iD)^2h_vi\int
d^3x\int_{-\infty}^0 \!\!\!\! dt \; {\cal
L}_I(x)| B(v)\rangle+h.c.&=&\frac{{\cal T}_1 + 3 {\cal
T}_2}{m_b}, \nonumber\\ 
\frac{1}{2M_B} \langle B(v)|\bar{h}_v
\frac{1}{2}(-i \sigma_{\mu\nu})G^{\mu\nu} h_vi\int d^3x\int_{-\infty}^0
  \!\!\!\! dt\; {\cal L}_I(x)| B(v)\rangle+h.c.&=&\frac{{\cal T}_3+3{\cal
T}_4}{m_b}
\end{eqnarray}
arising from a mismatch between the states $|B(v)\rangle$ of the effective
theory and $|B\rangle$ of the full theory.
The contributions from ${\cal T}_{1-4}$ can most easily be 
incorporated by making the replacements \cite{gremm}
\begin{eqnarray}
\lambda_1 &\to& \lambda_1 + \frac{{\cal T}_1 + 3 {\cal T}_2}{m_b}
\nonumber\\
\lambda_2 &\to& \lambda_2 + \frac{{\cal T}_3 + 3 {\cal T}_4}{3 m_b}
\label{tsubs}
\end{eqnarray}
in the parton level results.
In addition, as we will show later there is a contribution to the 
total rate from the dimension six four--fermion operator 
\begin{equation}
O_{(V-A)}^{bs} = 16 \pi^2 \; \left[ \bar{b} \gamma^\mu L s \bar{s} \gamma^\nu L b \;
                 ( g_{\mu\nu} - v_\mu v_\nu ) \right],
\end{equation}
the matrix element of which we define as
\begin{equation}
\label{f1definition}
\frac{1}{2 M_B} \langle B | O_{(V-A)}^{bs} | B \rangle \equiv f_1.
\end{equation}
The form factors up to ${\cal O}(1/m_b^2)$ have appeared in the literature 
\cite{ahhm_btosee_lepton}.
The ${\cal O}(1/m_b^3)$ contributions to the form factors proportional to $\rho_{1-2}$
are presented in Appendix \ref{formfactorAppendix}.  The dependence on
${\cal T}_{1-4}$ is obtained by making 
the replacements (\ref{tsubs}) in the ${\cal O}(1/m_b^2)$ form factors.

The triple differential branching ratio is given by 
\begin{eqnarray}
\label{triplelepdiff}
\frac{d^3{\cal B}}{d\hat{u}\,d\hat{s}\,dv\cdot\hat{q}} &=&
\frac{2 \;{\cal B}_0 }{2M_B} \left(-\frac{1}{\pi} \right) \mbox{Im}
\left\{ 2 \, \hat{s} \, \left(T_1^L(v\cdot \hat{q},\hat{s}) + T_1^R(v\cdot \hat{q},\hat{s}) \right) 
    \right. \nonumber \\ &+& \left( (v \cdot \hat{q})^2 - \hat{s} - 
\frac{\hat{u}^2}{4} \right) \left( T_2^L(v\cdot \hat{q},\hat{s}) +
T_2^R(v\cdot \hat{q},\hat{s}) \right) \nonumber \\ &+& \left. \hat{u} \, \hat{s} \,  
 \left( T_3^L(v\cdot \hat{q},\hat{s}) - T_3^R(v\cdot \hat{q},\hat{s}) \right)
\right\}, 
\end{eqnarray}
where we have defined kinematic variables 
$v\cdot \hat{q} = \frac{1}{m_b} v \cdot q$, $\hat{s} = \frac{1}{m_b^2} q^2$, 
and $\hat{u} = \frac{1}{m_b^2} \left[ (p_b - p_-)^2 - (p_b - p_+)^2 \right]$.
In terms of these leptonic variables the limits of phase space are
given by
\begin{eqnarray} 
\label{lepton_phase}
- \sqrt{\hat{s}+\frac{\hat{u}^2}{4}} &\le& v \cdot \hat{q} \le
\sqrt{\hat{s}+\frac{\hat{u}^2}{4}} \nonumber\\ 
-\hat{u}(\hat{s},\hat{m}_s)&\le& \hat{u} \le \hat{u}(\hat{s},\hat{m}_s)\nonumber\\ 
4 \hat{m}_l^2 &\le& \hat{s} \le \left(1-\hat{m}_s\right)^2,
\end{eqnarray} 
where $\hat{u}(\hat{s},\hat{m}_s) =
\sqrt{\left[\hat{s}-(1+\hat{m}_s)^2 \right]
  \left[\hat{s}-(1-\hat{m}_s)^2 \right]}$. 

For the calculation of the hadron invariant mass moments it will be
convenient to express the phase space in terms of 
the parton energy fraction $x_0 = E_q/m_b$ and the
parton invariant mass fraction 
$\hat{s}_0 = p_q^2/m_b^2$.  They are related to the leptonic variables
introduced above via  
\begin{eqnarray}  
\label{lephadtrans}
v \cdot \hat{q} &=& 1 - x_0 \nonumber \\ 
\hat{s} &=& 1 + \hat{s}_0 - 2 x_0. 
\end{eqnarray} 
The phase space can then be expressed as
\begin{eqnarray}
- 2 \sqrt{x_0^2-\hat{s}_0} &\le& \hat{u} \le 2 \sqrt{x_0^2-\hat{s}_0}\nonumber\\ 
\hat{m}_s^2 &\le& \hat{s}_0 \le x_0^2 \nonumber\\
\hat{m}_s &\le& x_0 \le \frac{1}{2}(1+\hat{m}_s^2).
\end{eqnarray}
Since the form factors $T_i$ are independent of $\hat{u}$, this first
integration is trivial and we arrive 
at 
\begin{eqnarray}
\label{doublehaddiff}
\frac{d^2{\cal B}}{dx_0 \, d\hat{s}_0} &=&  \frac{16 \;{\cal B}_0 }{2M_B}
\left(-\frac{1}{\pi} \right) \sqrt{x_0^2 - \hat{s}_0} \mbox{Im}
\Bigg\{ \Bigg[ (1 - 2 x_0 + \hat{s}_0)
    \left( T_1^L(x_0,\hat{s}_0) + T_1^R(x_0,\hat{s}_0) \right)
  \nonumber \\ &+& \left. \left. \frac{x_0^2 -
      \hat{s}_0}{3}  \left( T_2^L(x_0,\hat{s}_0) + T_2^R(x_0,\hat{s}_0) \right)
  \right] \right\}
\end{eqnarray} 
 
In the above expressions we use the same conventions as in
\cite{ahhm_btosee_lepton,ah_btosee_hadron} and normalize the $B \to
X_s \ell^+ \ell^-$ branching ratio to the semileptonic branching ratio 
\begin{equation}
d {\cal B}(B \to X_s \ell^+ \ell^-) = {\cal B}_{sl} \frac{d \Gamma(B
  \to X_s \ell^+ \ell^-)}{\Gamma(B \to X_c \ell \nu_{\ell})},
\end{equation}
introducing the normalization constant
\begin{equation}
{\cal B}_0 = {\cal B}_{sl} \frac{3 \alpha^2}{16 \pi^2} \frac{|V_{ts}^* 
  V_{tb}|^2}{|V_{cb}|^2} \frac{1}{f(\hat{m}_c) \kappa(\hat{m}_c)}.
\end{equation}
In this expression $f(\hat{m}_c)$ is the well-known phase space factor 
for the  $b \to c  e \bar{\nu}$ parton decay rate 
\begin{equation}
f(\hat{m}_c) = 1-8 \hat{m}_c^2 + 8 \hat{m}_c^6 - \hat{m}_c^8 - 24
\hat{m}_c^4 \log \hat{m}_c
\end{equation}
and $\kappa(\hat{m}_c)$ includes the ${\cal O}(\alpha_s)$ QCD
corrections as well as the nonperturbative corrections up to ${\cal
  O}(1/m_b^3)$
\begin{equation}
\kappa(\hat{m}_c) = 1 + \frac{\alpha_s(m_b)}{\pi} g(\hat{m}_c) +
\frac{h_1(\hat{m}_c)}{2 m_b^2} + \frac{h_2(\hat{m}_c)}{6 m_b^3}
\end{equation}
where 
\begin{eqnarray}
g(\hat{m}_c) &=& \frac{A_0(\hat{m}_c)}{f(\hat{m}_c)}\nonumber\\
h_1(\hat{m}_c) &=& \lambda_1 + \frac{\lambda_2}{f(\hat{m}_c)} \left(
  -9 + 24 \hat{m}_c^2 - 72 \hat{m}_c^4 + 72 \hat{m}_c^6 - 15
  \hat{m}_c^8 - 72 \hat{m}_c^4 \log \hat{m}_c \right) \nonumber\\
h_2(\hat{m}_c) &=& \frac{\rho_1}{f(\hat{m}_c)} 
 \left( 77 - 88 \hat{m}_c^2 + 24 \hat{m}_c^4 - 8 \hat{m}_c^6 - 5
   \hat{m}_c^8 + 96 \log \hat{m}_c + 72 \hat{m}_c^4 \log \hat{m}_c
 \right) \nonumber \\ 
 &+& \frac{\rho_2}{f(\hat{m}_c)} \left( 27 - 72 \hat{m}_c^2 + 216
   \hat{m}_c^4  - 216 \hat{m}_c^6 + 45 \hat{m}_c^8 + 216 \hat{m}_c^4 \log \hat{m}_c \right)  
\end{eqnarray} 
The analytic expression for the perturbative function $A_0(\hat{m}_c)$
can be found in \cite{hadmass}.

\section{The partially integrated branching ratio}
\label{section_partial}

An interesting experimentally accessible quantity is the
dilepton invariant mass spectrum.  
Evaluating the $\hat{u}$ integral in (\ref{triplelepdiff}), and doing
the integral over $v \cdot \hat{q}$
by picking out the residues, we find for the dilepton invariant mass
spectrum
\begin{eqnarray}
        \frac{{\rm d}{\cal B}}{{\rm d}\hat{s}} & = & 2 \; {\cal B}_0 
                \left\{ 
                  \left[
                \frac{1}{3} (1-\hat{s})^2 (1+2 \hat{s}) \; (2 +
                \frac{\lambda_1}{m_b^2})  
                + \left( 1 - 15  \hat{s}^2 + 10 \hat{s}^3\right)
                (\frac{\lambda_2}{m_b^2} - \frac{\rho_2}{m_b^3})
              \right.\right.\nonumber\\ 
        & &  \left.\left.\;\;\;\;\;\;\;\;\;\;\;\;\;\;
                -  \frac{10 \hat{s}^4 + 23 \hat{s}^3 - 9 \hat{s}^2 + 13
                  \hat{s} +11}{9(1-\hat{s})} \frac{\rho_1}{m_b^3}
                        \right] 
                \left( |C_9^{\mbox{eff}}(\hat{s}) |^2 + C_{10}^2 \right)
                \right.\nonumber \\
        & & \left.
             +  \left[
                 \frac{4}{3} (1-\hat{s})^2 (2+ \hat{s}) \; (2 +
                 \frac{\lambda_1}{m_b^2}) 
                + 4  \left( -6 -3  \hat{s} + 5 \hat{s}^3 \right)
                (\frac{\lambda_2}{m_b^2} - \frac{\rho_2}{m_b^3}) \right. \right.\nonumber\\
        & &  \left.\left.\;\;\;\;\;\;\;\;\;\;\;\;\;\;
                - \frac{4(5 \hat{s}^4 + 19 \hat{s}^3 + 9 \hat{s}^2 - 7 
                  \hat{s} + 22)}{9(1-\hat{s})}
                \frac{\rho_1}{m_b^3}\right]
              \frac{|C_7^{\mbox{eff}}|^2}{\hat{s}} 
                \right.\nonumber \\
        & & \left.
           +    \left[
                4 (1-\hat{s})^2 (2+ \frac{\lambda_1}{m_b^2})
               + 4  \left( -5 -6  \hat{s} + 7 \hat{s}^2 \right)
               (\frac{\lambda_2}{m_b^2} - \frac{\rho_2}{m_b^3})
             \right. \right. \nonumber\\ & &
           \left.\left.\;\;\;\;\;\;\;\;\;\;\;\;\;\; 
                + \frac{4 ( 3 \hat{s}^3 - 17 \hat{s}^2 + \hat{s} -
                  3)}{3(1-\hat{s})} \frac{\rho_1}{m_b^3} \right]
              Re(C_9^{\mbox{eff}}(\hat{s})) \, C_7^{\mbox{eff}}  
                \right. \nonumber\\
        & &  \left.
- \frac{16}{3} \frac{\rho_1}{m_b^3} \delta(1-\hat{s}) \left(C_{10}^2+
  \left(C_9^{\mbox{eff}} (\hat{s})+2
  C_7^{\mbox{eff}}\right)^2 
\right) \right\}\, . 
\label{eqn:dbds0}
\end{eqnarray}
The dependence on ${\cal T}_{1-4}$ can be obtained by making the
replacements (\ref{tsubs}) in (\ref{eqn:dbds0}).  In this expression we have
taken the limit $\hat{m}_s \to 0$.  The corresponding expression with full
$\hat{m}_s$ dependence is given in Appendix \ref{specwithmass}.

\begin{figure}[htbp]
\centerline{\epsfxsize=10 cm \epsfbox{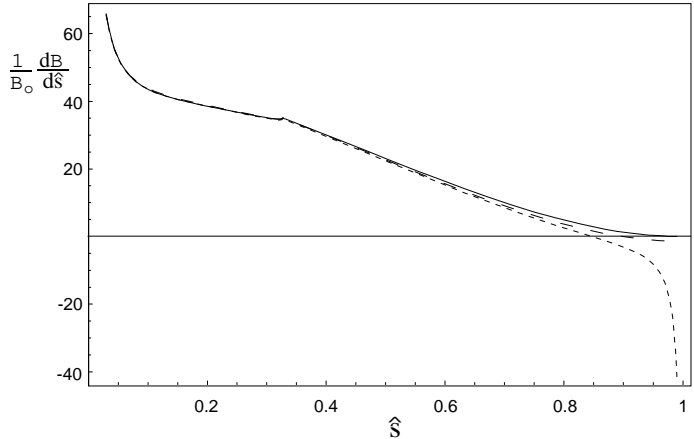}}
\caption{The differential decay spectrum. The solid line shows the
  parton model prediction, the long-dashed line includes the 
  ${\cal O}(1/m_b^2)$
  corrections and the short-dashed line contains all corrections up to $
  {\cal O}(1/m_b^3)$.} 
\label{diff_spec} 
\end{figure} 

A plot of this distribution is shown in Fig.~\ref{diff_spec}, where we
have used the values for the nonperturbative matrix elements 
\begin{equation} 
\label{parameter_set} 
\lambda_1 = -0.19 \; \mbox{GeV}^2,\;
 \lambda_2 = 0.12 \; \mbox{GeV}^2.
\end{equation}   
For the matrix elements of the
dimension six operators we use the generic size $(\lqcd)^3 \sim (0.5 \; 
{\rm GeV})^3$ as suggested by dimensional analysis. The vacuum saturation approximation \cite{vac_sat} predicts $\rho_1 > 0$, as shown, and we find the 
displayed spectrum is fairly insensitive to the sign of the other dimension
six matrix elements.  One immediately 
notices divergences at both endpoints of this
spectrum. The divergence at the $\hat{s} \to 0$ endpoint is due to the
intermediate photon going on--shell and is a well known feature of
the decay $B \to X_s \ell^+ \ell^-$ \cite{grinstein_savage_wise,ahhm_btosee_lepton}.  In 
this limit one expects this spectrum to reduce to the 
$B \to X_s \gamma$ rate with an on--shell
photon in the final state, convoluted with the fragmentation function giving
the probability for 
a photon to fragment into a lepton pair.  This correspondence is
explicitly verified by the 
analytic form of the divergent term 
\begin{equation}
\frac{1}{{\cal B}_0} \left. \frac{d{\cal B}}{d\hat{s}}
\right|_{\hat{s} \to 0} \sim \frac{32}{3} \frac{|C_7^{\mbox{eff}}|^2}{\hat{s}} \left( 1 +
  \frac{\lambda_1 - 9 \lambda_2}{2m_b^2} -
  \frac{11\rho_1 - 27 \rho_2}{6 m_b^3} + \frac{{\cal T}_1 + 3 {\cal T}_2 -
    3 \left({\cal T}_3 + 3 {\cal T}_4 \right)}{2m_b^3}\right)
\end{equation}
where the term multiplying $1/\hat{s}$ is proportional to the total
rate for $B \to X_s \gamma$  
\cite{christian}.  As mentioned above, experimental cuts require us to stay away
from this endpoint, and therefore automatically regulate this divergence.

The divergence at the $\hat{s} \to 1$ endpoint is entirely due to
the $1/m_b^3$ operators as can be seen from Fig.~\ref{diff_spec}.  In 
this case the analytic form of the divergent term is
\begin{equation}
\label{s1divergence}
\frac{1}{{\cal B}_0} \left. \frac{d{\cal B}}{d\hat{s}} \right|_{\hat{s} \to 1} \sim
\frac{32}{3 \, m_b^3} \left(C_{10}^2 + (2 \, C_7^{\mbox{eff}} + C_9^{\mbox{eff}}(\hat{s}) )^2\right)
\frac{\rho_1}{1 -\hat{s}}. 
\end{equation}
This leads, upon integration, to an unphysical 
logarithmic divergence in the expression for the total rate that is
regulated by the mass of the $s$ quark. (Of course, it is only consistent to 
include the mass of the $s$ quark in the upper limit of integration if one 
uses the spectrum with the full $m_s$ dependence as given in Appendix
\ref{specwithmass}.) 
This divergence can be understood by considering a similar effect in the semileptonic 
decay $B \to X_c \ell \bar{\nu}_\ell$ \cite{blok}. In that context,
the origin of this divergence can be clarified by performing an OPE 
for the total, rather than the differential, rate \cite{blok,sumlogs}.
Including dimension six operators in this OPE one finds a four
fermion operator of the form
\begin{equation}
\frac{16 \pi^2}{m_b^3}  \bar{b} \gamma^\mu L c \bar{c} \gamma^\nu L b \;
                 ( g_{\mu\nu} - v_\mu v_\nu )
\end{equation}
contributing to the rate.
In \cite{blok} the matrix element of this operator was calculated at leading order
in perturbation theory by
integrating out the $c$ quark and its contribution to the total rate
was found to be $\rho_1 \log(\hat{m}_c)$.
To calculate this matrix element it was essential 
that the mass of the $c$ quark be large compared to the QCD scale
$\Lambda_{QCD}$.  Consequently, for the decay $B \to X_s \ell^+ \ell^-$ where the same operator with $s$ quarks rather than $c$ quarks appears, these methods are not
applicable because the $s$ quark is too light. Including higher orders in perturbation theory
the matrix element of 
the four fermion operator contains logarithms of the form $\alpha_s^n
\log^{n+1}(\hat{m}_s)$ which are 
of order unity, making a perturbative calculation of this matrix
element impossible. Thus, a seventh non-perturbative
matrix element $f_1$ defined in (\ref{f1definition}) is required. 
It contributes only at the $\hat{s} \to 1$ endpoint of 
the spectrum and cancels the logarithmic divergence proportional to
$\rho_1 \log(\hat{m}_s)$ in the total rate
\begin{equation}
\frac{d{\cal B}}{d\hat{s}} \to \frac{d{\cal B}}{d\hat{s}} - 
 \frac{32}{3 m_b^3} {\cal B}_0 \left(C_{10}^2 + (2 \, C_7^{\mbox{eff}} + C_9^{\mbox{eff}}(\hat{s}) )^2\right) \,\,\delta\left( 1-\hat{s}
 \right) \left(\rho_1 \log(\hat{m}_s) - f_1
  \right).
\end{equation}

Another noticeable feature in the dilepton invariant mass spectrum is the 
cusp due to the $c \bar{c}$ threshold. Near this value of $\hat{s}$ the 
methods we used to
calculate the physical spectrum fail because of long distance contributions
from the resonant decay $B \to X_s J/\psi$, where the $J/\psi$ subsequently decays
into two leptons. 
Experimentally one deals with this resonance region
by simply cutting it out. Thus, to compare reliably to experiment we
should include such a cut in our calculation. Defining the partially
integrated branching ratio by
\begin{equation}
\label{bchi} 
{\cal B}_\chi = \frac{1}{{\cal B}_0} \int_\chi^1 \, d\hat{s} 
  \frac{d {\cal B}}{d\hat{s}} 
\end{equation} 
\begin{figure}[htbp]
\centerline{\epsfxsize=10 cm \epsfbox{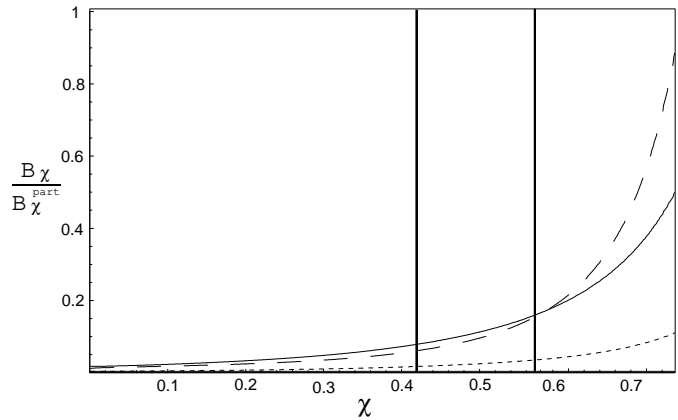}}
\caption{The fractional contributions to $B_\chi$ with respect to
  the parton model result from the ${\cal O}(1/m_b^3)$ operators. The
  solid, dashed and dotted lines correspond to the contributions from
  $\lambda_2$, $\rho_1$ and $\rho_2$, respectively. The contribution
 from $\lambda_1$ is too small to be seen. The two vertical lines
 illustrate the positions of the $J/\psi$ and the $\psi'$ resonance.}
\label{cut_plot} 
\end{figure}
we plot the contribution of
the individual matrix elements relative to the leading order parton
result in Fig.~\ref{cut_plot}.  For the generic size $\rho_i \sim (\lqcd^3)$ used
in this plot, the contribution from $\rho_1$ is of the same
size as the contribution from dimension five operators. 
This implies that 
including the ${\cal O}(1/m_b^2)$ corrections for this decay does not 
significantly decrease the 
nonperturbative uncertainties.  We see that the nonperturbative contributions 
become more dominant as the
accessible phase space is decreased\footnote{We emphasize that the sizes of the $\rho_i$ 
contributions shown here should not be taken as accurate indications
of the actual size of the corrections, but rather as estimates of the
uncertainty in the prediction.}. 
For $\chi \sim .75$ the 
uncertainty from the $\rho_1$ matrix element is of the same size as
the parton model prediction. This is a clear signal that the OPE is no 
longer valid if the phase space is restricted to be too close to the
endpoint $\hat{s} = 1$. This breakdown of the OPE close to the
endpoint is a well--known feature encountered in this approach to 
the study of inclusive decays \cite{shape_func}. Unfortunately, in this endpoint region a
shape function does not exist and an 
alternate approach, such as heavy hadron chiral perturbation theory for exclusive final states, 
must be used \cite{OPEbreakdown}.

A cut of $\chi = (14.33\;\mbox{GeV}^2/m_b^2) = 0.59$ has been suggested by the 
CLEO collaboration in order to eliminate the resonance region \cite{experimental_search}.
For this value the partially integrated rate is
\begin{eqnarray}
\label{CLEObranchingratio}
  {\cal B}_{0.59} &=& 3.8 +
  1.9 \left(\frac{\lambda_1}{m_b^2} + \frac{{\cal T}_1 + 3 {\cal
        T}_2}{m_b^3}\right) - 134.7 \left( \frac{\lambda_2}{m_b^2} +
    \frac{{\cal T}_1 + 3 {\cal
        T}_2}{3 m_b^3}\right) \nonumber\\
&& + 614.9 \frac{\rho_1}{m_b^3} 
        + 134.7 \frac{\rho_2}{m_b^3} + 560.2 \frac{f_1}{m_b^3}.
\end{eqnarray}
At this value of the cut $\chi$ the coefficients of the nonperturbative matrix 
elements clearly indicate a poorly converging OPE.  One can estimate the uncertainty
induced by the ${\cal O}(1/m_b^3)$ parameters by fixing $\lambda_i$ to
the values given in (\ref{parameter_set}), 
then randomly varying the magnitudes of the parameters
$\rho_i, {\cal T}_i$ and $f_1$ between $-(0.5 \, \rm{GeV})^3$ and $(0.5 \,\rm{GeV})^3$
as suggested by dimensional analysis.  We also impose positivity
of $\rho_1$ as indicated by the vacuum saturation approximation 
\cite{vac_sat}, and we enforce the constraint 
\begin{equation}
\rho_2 - {\cal T}_2 - {\cal T}_4 = 
    \frac{ \left( \frac{\alpha_s(m_c)}{\alpha_s(m_b)}
    \right)^{3/\beta_0} M_B^2 \Delta M_B (M_D + \bar{\Lambda}) - 
         M_D^2 \Delta M_D (M_B + \bar{\Lambda})}{M_B + \bar{\Lambda} - 
   \left( \frac{\alpha_s(m_c)}{\alpha_s(m_b)} \right)^{3/\beta_0}
   (M_D + \bar{\Lambda})}
\end{equation}
derived from the ground state meson mass splittings 
$\Delta M_H = M_{H^*} - M_H \;\; (H = B,D)$ \cite{gremm}.  Here $\beta_0$ is the
usual coefficient of the beta function $\beta_0 = 11-2/3 n_f =
25/3$ for $4$ light flavours. Taking the 1 $\sigma$ 
deviation as a reasonable estimate of the uncertainties from ${\cal O}(1/m_b^3)$
contributions, we find the
uncertainty in ${\cal B}_{0.59}$ to be at the 10\% level. It is clear from 
(\ref{CLEObranchingratio}) that the $\rho_1$ contribution is large, and 
relaxing the positivity constraint on $\rho_1$ enlarges the
uncertainty to about 20\%. A similar statement can be made 
regarding the $D\mbox{\O}$ analysis \cite{experimental_search} where the phase space cut is slightly higher, and the nonperturbative corrections are correspondingly somewhat larger.  
Since the cut on $\hat{s}$ cannot be lowered
because of the $\psi'$ resonance, these uncertainties are intrinsic
to our approach in the large dilepton invariant mass region.  

It is 
important to notice that in the invariant mass region below the
$J/\psi$ 
resonance, the uncertainties from these matrix elements are much
smaller.  
For example, integrating the differential spectrum up to the cut
specified 
in the CLEO analysis \cite{CLEOex} 
$ \hat{s} = (M_{J/\psi} - 0.1 GeV)^2/m_b^2 = 0.35$ we find 
\begin{equation} 
\int_{0.01}^{0.35}\,d\hat{s} \frac{d{\cal B}}{d\hat{s}} = 22.0 \left(
  1 + 0.5 \right( \frac{\lambda_1}{m_b^2} + \frac{ {\cal T}_1 + 3
  {\cal T}_2 }{m_b^3} \left) + 1.2 \right( \frac{\lambda_2}{m_b^2} +
\frac{ {\cal T}_2 + 3 {\cal T}_4 }{3m_b^3} \left) - 3.7
  \frac{\rho_1}{m_b^3} - 1.2 \frac{\rho_2}{m_b^3} \right).
\end{equation}     
It is still true that the coefficient of the $\rho_1$ term
is $\sim 10$ times larger than that of the $\lambda_1$ term, but both sets of 
nonperturbative corrections are small relative to the parton level result in this region.
Although this does not allow us to draw a strong conclusion about the convergence of 
the OPE, we can conclude that in this region the ${\cal O}(1/m_b^3)$ nonperturbative corrections 
are not a significant source of theoretical uncertainty.  

\section{The Forward-Backward Asymmetry} 
\label{section_fbasym} 

The differential forward-backward asymmetry is defined by 
\begin{equation} 
\label{fbaDefinition} 
\frac{d {\cal A}}{d\hat{s}} = \int_0^1 \; dz \;
\frac{d\cal{B}}{dz\,d\hat{s}} 
  - \int_{-1}^0 dz \; \frac{d\cal{B}}{dz\,d\hat{s}} 
\end{equation} 
where  
\begin{equation} 
\label{zsub} 
z = cos\theta = \frac{\hat{u}}{\hat{u}(\hat{s},\hat{m}_s)} 
\end{equation} 
parameterizes the angle between the $b$ quark and the  
$\ell^+$ in the dilepton CM frame.  It has been shown
\cite{newphysics} 
that new physics can modify this spectrum, so it is  
interesting to see how ${\cal O}(1/m_b^3)$ terms contribute to the SM 
prediction.  

Integrating the triple differential decay rate (\ref{triplelepdiff}) we find 
\begin{eqnarray} 
\frac{d{\cal A}}{d\hat{s}} &=& C_7^{\mbox{eff}} C_{10} \left( -8 
        {{\left( 1 - \hat{s} \right) }^2}-  
     {\frac{4\left( 3 + 2\hat{s} + 3{{\hat{s}}^2} \right)
         \lambda_1}{3{m_b^2}}} +  
     {\frac{4 \left( 7 + 10\hat{s} - 9{{\hat{s}}^2} \right) 
         \lambda_2}{{m_b^2}}} \right.\nonumber\\  
&&\left.\qquad\qquad\qquad 
  +  
     {\frac{4\left( 5 + 2\hat{s} + {{\hat{s}}^2} \right)
         \rho_1}{3{m_b^3}}} -  
     {\frac{4\left( 7 + 10\hat{s} - 9{{\hat{s}}^2} \right) 
         \rho_2}{{m_b^3}}} \right) \nonumber \\  
&+&   
   C_{9}^{\mbox{eff}}(\hat{s}) C_{10}\left( -4\hat{s} 
         {{\left( 1 - \hat{s} \right) }^2}  
            - {\frac{2\hat{s}\left( 3 + 2\hat{s} + 
             3{{\hat{s}}^2} \right) \lambda_1} 
       {3{m_b^2}}} + {\frac{2\hat{s}\left( 9 + 14\hat{s} - 
           15{{\hat{s}}^2} \right) \lambda_2} 
       {{m_b^2}}} \right.\nonumber\\ 
&&\left.\qquad\qquad\qquad 
  - {\frac{2\hat{s}\left( 1 + 2\hat{s} + 5{{\hat{s}}^2} \right)
      \rho_1} 
       {3{m_b^3}}} - {\frac{2\hat{s}\left( 1 + 6\hat{s} - 
           15{{\hat{s}}^2} \right) \rho_2} {{m_b^3}}} \right) 
\end{eqnarray} 
Here we have again omitted the trivial dependence on ${\cal T}_{1-4}$ . 

It is clear from this expression that the third order 
terms do not have 
abnormally large coefficients, and therefore introduce only small  
variations relative to the second order expression.   
 
\begin{figure}[htbp] 
\centerline{\epsfxsize=10 cm \epsfbox{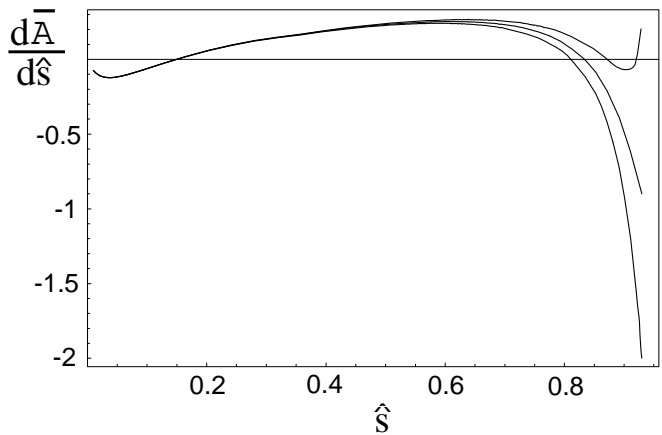}}
\caption{The normalized forward backward asymmetry.  The three curves show 
the mean value and the $1\sigma$ uncertainty of the forward 
backward asymmetry, obtained in a way similar to that explained in
section 
\ref{section_partial}. } 
\label{asymm}  
\end{figure}

An experimentally more useful quantity is the normalized FB asymmetry
defined 
by 
\begin{equation} 
\frac{d \bar{{\cal A}}}{d\hat{s}} = \frac{d{\cal
    A}}{d\hat{s}}/\frac{d{\cal B}}{d\hat{s}}. 
\end{equation} 
Unfortunately, this normalized asymmetry  
has inherited the poor behavior of the differential 
branching ratio in the endpoint region. In Fig. \ref{asymm}  
we illustrate  
the uncertainties of the normalized FB asymmetry originating from the 
matrix elements of the dimension six operators. The three curves show 
the mean value and the $1 \sigma$ uncertainty of the forward 
backward asymmetry, obtained in a way similar to that explained in
section 
\ref{section_partial}. We can see that up to a value of $\hat{s} =
0.7$ 
the uncertainties are small, but for larger values of the dilepton 
invariant mass the uncertainties increase rapidly. Because of the 
necessity of the cut to eliminate the $c \bar{c}$ resonances, the 
accessible high dilepton invariant mass region is therefore restricted
to a few
hundred MeV. For the dilepton invariant mass region below the $J/\psi$ 
resonance $(\hat{s} < 0.35)$ the uncertainties are small.  
 
\section{Extracting $\bar{\Lambda}$ and $\lambda_1$ from the hadron
  invariant mass moments}
\label{section_moments} 
Throughout this paper we have fixed the values of the leading
non-perturbative  
parameters $\bar{\Lambda}$, $\lambda_1$, and $\lambda_2$.  However,
these 
values must be determined from experiment.  The most sensitive
observables 
for this purpose are those which vanish in the parton model.  It is
interesting to 
ask how severely our ignorance of the values of the ${\cal O}(1/m_b^3)$ 
parameters compromises our ability to extract the values of $\bar{\Lambda}$
and $\lambda_1$ from such a measurement.  It has been suggested by 
Ali and Hiller \cite{ah_btosee_hadron} that one use the first
two moments of the hadron invariant mass spectrum defined by
\begin{equation}
\label{had_moments}
\langle S_H^n \rangle = \int (S_H - M_H^2)^n \frac{d{\cal B}}{dS_H} dS_H.
\end{equation}
This idea
is similar to the approaches used in the semileptonic $B\to X_c \ell \nu_\ell$
 \cite{gremm} and the rare radiative $B \to X_s \gamma$ decays 
\cite{christian}, though the experimental task is more difficult in this
case due to the small size of the branching ratio.
To calculate these
hadronic moments we relate them to calculable partonic moments
via
\begin{eqnarray}\label{sHmoments}
\langle S_H \rangle &=& \bar{\Lambda}^2 -
\frac{\bar{\Lambda}(\lambda_1 + 3 \lambda_2)}{M_B} \nonumber\\
&&+ \left( M_B^2 - 2
  M_B \bar{\Lambda} + 
  \bar{\Lambda}^2 + \lambda_1 + 3 \lambda_2 -\frac{\rho_1 + 3
    \rho_2}{2M_B} + \frac{ {\cal T}_1 + {\cal T}_3 + 3 ( {\cal T}_2 +
    {\cal T}_4 )} {2M_B} \right)  \langle \hat{s}_0 \rangle \nonumber\\
&& + \left( 2 M_B \bar{\Lambda} - 2 \bar{\Lambda}^2 - \lambda_1 -3
  \lambda_2 + \frac{\bar{\Lambda}(\lambda_1 + 3 \lambda_2)}{M_B} +
  \frac{\rho_1 + 3 \rho_2}{2 M_B} \right.\nonumber\\
&&\qquad\qquad\qquad \left.
  - \frac{ {\cal T}_1 + {\cal T}_3 + 3
    ( {\cal T}_2 + {\cal T}_4 )}{2M_B}\right) \langle x_0 \rangle \\
\langle S_H^2 \rangle &=& \left( M_B^4 - 4 M_B^3 \bar{\Lambda} + 6
  M_B^2 \bar{\Lambda}^2 + 2 M_B^2 (\lambda_1 + 3 \lambda_2) - 4 M_B
  \bar{\Lambda}^3 - 4 M_B \bar{\Lambda}(\lambda_1 + 3
  \lambda_2) \right. \nonumber\\
&& \left. \qquad \qquad \qquad - M_B 
  (\rho_1 + 3 \rho_2) + M_B ({\cal T}_1 + {\cal T}_3 + 3 ( {\cal T}_2
  + {\cal T}_4 ) ) \right) \langle \hat{s}_0^2 \rangle \nonumber\\
&& + 4 \left( M_B^2 \bar{\Lambda}^2 - 2 M_B \bar{\Lambda}^3 - M_B
  \bar{\Lambda}(\lambda_1 + 3 \lambda_2) \right) \langle x_0^2 \rangle 
\nonumber\\
&& + \left( 4 M_B^3 \bar{\Lambda} - 12 M_B^2 \bar{\Lambda}^2 - 2 M_B^2 
  (\lambda_1 + 3 \lambda_2) + 12 M_B \bar{\Lambda}^3
+ 10 M_B
  \bar{\Lambda} (\lambda_1 + 3 \lambda_2) \right. \nonumber\\
&& \left. \qquad \qquad \qquad + M_B(\rho_1 + 3 \rho_2) - 
  M_B ({\cal T}_1 + {\cal T}_3 + 3 ( {\cal T}_2 + {\cal T}_4 ) )  
\right) \langle x_0 \hat{s}_0 \rangle \nonumber\\
&& + 2 \left(  M_B^2 \bar{\Lambda}^2 - 2 M_B \bar{\Lambda}^3 -
  (\lambda_1 + 3  
  \lambda_2) M_B \bar{\Lambda} \right) \langle \hat{s}_0 \rangle \nonumber \\
&& + 4 M_B \bar{\Lambda}^3 \langle x_0 \rangle
\end{eqnarray}
where we have used the mass relation 
\begin{equation}
\label{mass_relation} 
m_b = M_B - \bar{\Lambda} + \frac{\lambda_1 + 3 \lambda_2}{2 m_b} -
\frac{\rho_1 + 3 \rho_2}{4 m_b^2} + \frac{ {\cal T}_1 + {\cal T}_3 + 3
  ( {\cal  T}_2 + {\cal T}_4 )}{4 m_b^2}
\end{equation}
appropriate at this order in the OPE.  
We therefore have to calculate the first two moments of the parton
energy $\langle x_0 \rangle,\langle x_0^2 \rangle$ and parton invariant mass
$\langle \hat{s}_0 \rangle,\langle \hat{s}_0^2 \rangle$, as well as
the mixed moment $\langle x_0 \hat{s}_0 \rangle$.
Defining
\begin{equation}
M^{(m,n)} = \langle x_0^m \hat{s}_0^n \rangle = 
\frac{1}{{\cal B}_0} \int_{\hat{m}_s}^{\frac{1}{2}(1-\chi)} dx_0 \int_{\hat{m}_s^2}^{x_0^2} 
d\hat{s}_0 \; x_0^m
\hat{s}_0^n \frac{d^2 {\cal B}}{dx_0 d\hat{s}_0},
\end{equation}
we give the results for the required partonic moments in Appendix
\ref{parton_moments}. As before, we have included the dependence on
the cut on  the 
lepton invariant mass in these results. It is important to note that the
results for the partonic moments given in Appendix \ref{parton_moments} are
expressed in terms of the $b$ quark mass $m_b$ and must be
re-expressed in terms of the $B$ meson mass $M_B$ using the mass relation
(\ref{mass_relation}). Using again for the cut on the
invariant mass  the value proposed by CLEO $q^2 > 14.33 \;
\rm{GeV}^2$, we find for the two moments 
\begin{eqnarray}
\langle S_H \rangle &=& M_B^2 \left[ 0.36 \frac{\bar{\Lambda}}{M_B} +
  0.64 \frac{\lambda_1}{M_B^2} + 0.67 \frac{\lambda_2}{M_B^2} - 0.09
  \frac{\bar{\Lambda}^2}{M_B^2} + 8.48 \frac{\rho_1}{M_B^3} + 3.79
  \frac{\rho_2}{M_B^3}  \right. \nonumber \\ 
&&\qquad\qquad \left. + 1.09
  \frac{ \bar{\Lambda} \lambda_1}{M_B^3}  + 4.88 \frac{ \bar{\Lambda}
    \lambda_2 }{M_B^3} - 0.41 \frac{\bar{\Lambda}^3}{M_B^3} + 0.73
  \frac{{\cal T}_1 + 3 {\cal T}_2}{M_B^3} + 0.31 \frac{{\cal T}_3 + 3
    {\cal T}_4}{M_B^3} \right]\\ 
\langle S_H^2 \rangle &=& M_B^4 \left[ - 0.05 \frac{\lambda_1}{M_B^2}
  + 0.14 \frac{\bar{\Lambda}^2}{M_B^2} - 0.53 \frac{\rho_1}{M_B^3} -
  0.21 \frac{\rho_2}{M_B^3} + 0.63 \frac{ \bar{\Lambda}
    \lambda_1}{M_B^3}  + 0.46 \frac{ \bar{\Lambda} \lambda_2 }{M_B^3}
  \right. \nonumber \\ 
&&\qquad\qquad \left. - 0.05 \frac{{\cal T}_1 + 3 {\cal T}_2}{M_B^3} \right]
\end{eqnarray}
Consider first the expression for $\langle S_H^2 \rangle$.  The $\lambda_1$ 
term has a small coefficient and tends to cancel against higher order 
corrections, making this moment particularly insensitive to $\lambda_1$.  
We can see the problem another way by solving
this equation for $\lambda_1$: the solution exhibits a pole near
$\bar{\Lambda} = 0.4$, close to the expected value of $\bar{\Lambda}$ \cite{lam1values}.  As a result, the extracted value of $\lambda_1$ is extremely sensitive to the values of the higher order parameters.
Since the presence of this pole persists as the value of the cut is
changed, we conclude that this observable is unsuitable for extracting $\lambda_1$.   

\begin{figure}[htbp]
\centerline{\epsfxsize=10 cm \epsfbox{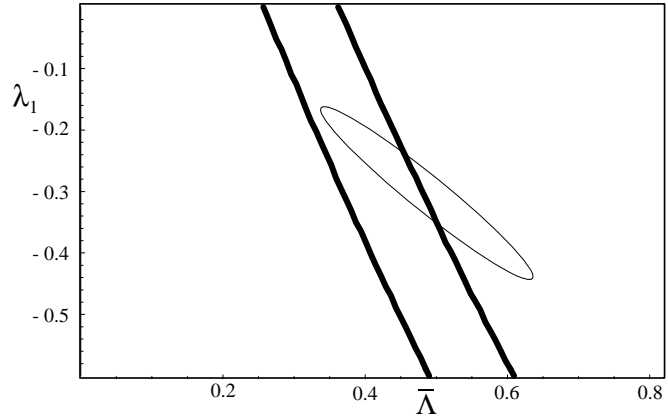}}
\caption{The constraint in the $\bar{\Lambda}-\lambda_1$ plane from $\langle S_H \rangle$. The width of the band is entirely
due to uncertainties from ${\cal O}(1/m_b^3)$ operators.  The ellipse is the equivalent constraint from moments of semileptonic $B \to X_c \ell \nu_\ell$.  It is only the relative orientation, and not location, of the constraints which has meaning.} 
\label{band}
\end{figure}

For the first moment $\langle S_H \rangle$ the convergence of the OPE is much
better.  Estimating the uncertainties from the unknown values of the 
dimension six operators by the method explained in section
\ref{section_partial}, 
we present the resulting constraint in the $\bar{\Lambda}-\lambda_1$ plane 
in Fig.~\ref{band}.
Superimposed on this figure is the ellipse obtained in
\cite{leptoncut} from an analogous study of moments of $B \to X_c \ell \nu_\ell$.  
Unfortunately, the bound from our analysis is nearly parallel to the major axis of 
this ellipse and, since it is only the relative orientation of the constraints 
which has meaning in this figure, 
this moment does not provide much additional information about
the values of $\bar{\Lambda}$ or $\lambda_1$.

\section{Conclusions}

Our purpose in this paper has been to study nonperturbative uncertainties 
in the rare inclusive decay $B \to X_s \ell^+ \ell^-$.  Building on previous
studies which evaluated the leading nonperturbative corrections, we have
parameterized the corrections arising at ${\cal O}(1/m_b^3)$ 
in terms of two matrix elements of 
local operators $\rho_{1,2}$, four matrix elements of non local
operators ${\cal T}_{1-4}$, and one matrix element of a four fermion
operator $f_1$. 

The numerical values of these parameters are unknown, yet even so 
a knowledge of the analytic form of the corrections allows us to 
study the convergence properties of the operator product expansion in
various regions of phase space.  Furthermore, the assumption that
these parameters, being of nonperturbative origin, should be ${\cal O}(\lqcd^3)$
permits us to make numerical estimates of theoretical uncertainties in
observable quantities.

We first considered the corrections to the differential spectrum 
$d{\cal B}/d\hat{s}$.  The experimental spectrum contains two 
prominent resonances due to intermediate $J/\psi$ and $\psi'$ production,
and the necessity of cutting these resonances out divides the accessible
spectrum into two parts: the region of low dilepton invariant mass 
below the $J/\psi$ resonance, and the region of high dilepton invariant
mass above the $\psi'$ resonance.  In the first region, we find that the
parton level calculation dominates, and that nonperturbative corrections
are small.  The operator product expansion appears to be converging 
according to expectation and, should it be possible to take experimental
data in this region, the results will not suffer from significant 
nonperturbative uncertainties.  

On the other hand we do find that, as expected, the nonperturbative uncertainties
increase as one moves into the high dilepton invariant mass region. It is
well known that as one approaches this endpoint, 
the operator product expansion breaks down.  The interesting result of
our study, however, is that the expansion breaks down somewhat earlier than was 
anticipated, with the ${\cal O}(1/m_b^3)$ uncertainties coming to
dominate the integrated rate once the available range of $\hat{s}$ is reduced
to about one quarter of the full range.  We also showed that the rate
obtained by integrating over the entire region above the $\psi'$ resonance
contains uncertainties from dimension six matrix elements at the $10\%$ level,
the uncertainties being dominated by those from the $\rho_1$ matrix element.  
This result may impact the potential for doing precise searches for
new physics using data from this region of phase space.

We also studied the contributions from dimension six operators to the 
forward-backward asymmetry.  This quantity probes different combinations
of Wilson coefficients than the rate, and has been proposed as a 
complimentary source of information about possible new physics effects. As
with the rate, the spectrum $d{\cal A}/d\hat{s}$ is divided into two regions
by the $J/\psi$ and $\psi'$ resonances.  We find that the dimension six
contributions are not unduly large anywhere in the phase space, suggesting
that this observable has a well behaved OPE.  However
if the differential forward-backward asymmetry is normalized to the 
differential rate, the resulting spectrum contains large uncertainties 
in the high dilepton invariant mass region.  In the region of phase space below
the $J/\psi$ resonance, however, we find the nonperturbative corrections to
be small.

Finally we addressed a recent proposal suggesting that hadron invariant mass
moments of the differential spectrum for $B \to X_s \ell^+ \ell^-$ could,
due to the sensitivity of these moments to nonperturbative effects, be used
to constrain the values of the HQET parameters $\lambda_1$ and 
$\bar{\Lambda}$.  In the low $S_H$ region the moments
are suppressed relative to the rate and, considering the already tiny
branching ratio, it is unlikely that experimental measurements in this region
will be forthcoming.  Therefore we focused our attention on the high invariant
mass region.  In this region, we found that the first of these 
moments $\langle S_H \rangle$ provided a constraint in the 
$\bar{\Lambda}-\lambda_1$ plane, but that this 
constraint was nearly the same as those derived from other, more 
experimentally promising, processes, and therefore seems to be
of limited interest for this purpose.  As for the second invariant mass moment
$\langle S_H^2 \rangle$, we found that the nonperturbative uncertainties were such 
that it was not possible to extract a stable constraint on the values
of $\bar{\Lambda}$ or $\lambda_1$.  From these results, we conclude that
these moments are not well suited to the extraction of these parameters.

\section{acknowledgements}

We would like to thank Michael Luke for helpful discussions.  This work
was supported in part by the National Science and Engineering Research
Council of Canada.

\appendix
\section{The contribution of dimension six operators to the form factors}
\label{formfactorAppendix}

 In this appendix we present the form factors $T_i$.  
These form factors have been 
calculated previously up to ${\cal O}(1/m_b^2)$
\cite{ahhm_btosee_lepton}, and  we do not reproduce those results
here.  We decompose the new contributions arising  at ${\cal
  O}(1/m_b^3)$ as  
\begin{equation}
T_i^{{L/R}} = 2M_B \left(T_i^{(C_9\pm C_{10})^2} \left(C_{10}\pm
    C_9^{\mbox{eff}}\right)^2 +  
  T_i^{C_7^2} |C_7^{{\mbox{eff}}}|^2 + T_i^{C_7 (C_9\pm C_{10})} C_7^{eff} (C_9
  \pm C_{10})  
\right).
\end{equation}
For completeness we have included the full $\hat{m}_s$ dependence in
these expressions, though in our analysis we set $\hat{m}_s = 0$.
Defining $ x = 1 - 2 v \cdot \hat{q} + \hat{s} + i \epsilon$ with
$\hat{q} = \frac{q}{m_b}$, $\hat{s} = \hat{q}^2$ and $\hat{\rho_i} =
\rho_i/m_b^3$, ,  we find that the
third order contributions are 
\begin{eqnarray}
T_1^{(C_9\pm C_{10})^2} &=& {\frac{1}{12x}}(\hat{\rho}_1 + 3\hat{\rho}_2)
\nonumber\\ 
&&-
{\frac{1}{6{x^2}}} \left( \left( 2 + \hat{s} \right) 
  \hat{\rho}_1 - 3
  \left( 2 - \hat{s} \right) \hat{\rho}_2 + 
  \left( \hat{\rho}_1 + 3\hat{\rho}_2 \right) 
  v\cdot \hat{q} - 
  \left( \hat{\rho}_1 + 3\hat{\rho}_2 \right) 
  {{\left( v\cdot \hat{q} \right) }^2}
\right) \nonumber \\ 
&&- {\frac{2}{3{x^3}}} \left( \hat{\rho}_1 + 3\hat{\rho}_2 \right) 
     \left( 1 - v\cdot \hat{q} \right) 
     \left( \hat{s} - 
       {{\left( v\cdot \hat{q} \right) }^2}
        \right)  \nonumber\\
&& + {\frac{4}{3{x^4}}} \hat{\rho}_1{{\left( 1 - 
          v\cdot \hat{q} \right) }^2}
     \left( \hat{s} - 
       {{\left( v\cdot \hat{q} \right) }^2}
     \right) \\
T_2^{(C_9\pm C_{10})^2}  &=& -{\frac{1}{6\,x}}\left( \hat{\rho}_1 +
  3\hat{\rho}_2 \right)  \nonumber\\
&& -
{\frac{1}{3\,{x^2}}} \left( 4\hat{\rho}_1 + 6\hat{\rho}_2 - \left(
    \hat{\rho}_1 + 
    3\hat{\rho}_2 \right) v\cdot \hat{q} \right) \nonumber \\ 
&&+ {\frac{2}{3\,
    {x^3}}} \left( 3\hat{s}\hat{\rho}_2 - 2\left( 2\hat{\rho}_1 +
    3\hat{\rho}_2 \right) v\cdot \hat{q} +  
   2 \left( 2\hat{\rho}_1 + 3\hat{\rho}_2 \right) {{\left( v\cdot \hat{q}
           \right) }^2} \right) \nonumber\\
&&+ {\frac{8}{3\,{x^4}}}
     \hat{\rho}_1\left( 1 - v\cdot \hat{q} \right)  
     \left( \hat{s} - {{\left( v\cdot \hat{q} \right) }^2} \right) \\
T_3^{(C_9\pm C_{10})^2}  &=& -
{\frac{1}{6{x^2}}} \left( \hat{\rho}_1 + 3\hat{\rho}_2
          \right) v\cdot \hat{q} \nonumber \\ 
&&+ {\frac{2}{3\,
     {x^3}}} \left( 1 - v\cdot \hat{q} \right) 
     \left( 3\hat{\rho}_2 - 
       \left( \hat{\rho}_1 + 3\hat{\rho}_2 \right) 
        v\cdot \hat{q} \right) \nonumber \\ 
&&+ {\frac{4}{3\,{x^4}}} \hat{\rho}_1\left( 1 - 
       v\cdot \hat{q} \right) 
     \left( \hat{s} - 
       {{\left( v\cdot \hat{q} \right) }^2}
        \right) \\
T_1^{C_7^2} &=& {\frac{1}{3
     {{\hat{s}}^2}x}} \left( \left( \hat{\rho}_1 + 3\hat{\rho}_2
          \right) 
       \left( \left( 1 - 
            5{{\hat{m}_s}^2} \right) 
          \hat{s} - 
         2\left( 1 + {{\hat{m}_s}^2} \right)
            {{\left( v\cdot \hat{q} \right) }^
            2} \right)  \right) \nonumber \\ 
&&- {\frac{2}{3{{\hat{s}}^2}
     {x^2}}} \left( \hat{s}
        \left( \left( 
             4{{\hat{m}_s}^2} + \hat{s} + 
             {{\hat{m}_s}^2}\hat{s} - 4 \right)
           \hat{\rho}_1 + 
          3\left( 8{{\hat{m}_s}^2} + 
             \hat{s} + {{\hat{m}_s}^2}\hat{s}
              \right) \hat{\rho}_2 \right) \right.  \nonumber\\
&&\left.\;\;\;\;\;\;\;\;\;\;\;+ 
       \left( 1-5{{\hat{m}_s}^2} \right)
          \hat{s}
        \left( \hat{\rho}_1 + 3\hat{\rho}_2 \right) 
        v\cdot \hat{q} +
       \left( 1 + {{\hat{m}_s}^2} \right) 
        \left( \left( 8 - \hat{s} \right) 
           \hat{\rho}_1 - 3\hat{s}\hat{\rho}_2 \right)
          {{\left( v\cdot \hat{q} \right) }^2} \right.\nonumber\\
&&\left.\;\;\;\;\;\;\;\;\;\;\;- 2\left( 1 + {{\hat{m}_s}^2}
          \right) 
        \left( \hat{\rho}_1 + 3\hat{\rho}_2 \right) 
        {{\left( v\cdot \hat{q} \right) }^3}
        \right) \nonumber\\
&&- {\frac{8}{3{{\hat{s}}^2}
     {x^3}}} \left( 1 + {{\hat{m}_s}^2} \right) 
     \left( 1 - v\cdot \hat{q} \right) 
     \left( \hat{s}
        \left( \hat{\rho}_1 + 3\hat{\rho}_2 \right)  - 
       2\left( 2\hat{\rho}_1 + 3\hat{\rho}_2 \right)
          v\cdot \hat{q} \right) 
     \left( \hat{s} - 
       {{\left( v\cdot \hat{q} \right) }^2}
        \right) \nonumber\\
&&- {\frac{16}{3{{\hat{s}}^2}
     {x^4}}} \hat{\rho}_1
     \left( 1 - v\cdot \hat{q} \right) 
     \left( \hat{s} - 
       {{\left( v\cdot \hat{q} \right) }^2}
        \right) \nonumber\\
&&\;\;\;\;\;\;\;\;\;\;\;\;\;\;\;\;\;\left( \hat{s} + 
       3{{\hat{m}_s}^2}\hat{s} + 
       \left( 1 + {{\hat{m}_s}^2} \right) 
        \hat{s}\,v\cdot \hat{q} - 
       2\left( 1 + {{\hat{m}_s}^2} \right) 
        {{\left( v\cdot \hat{q} \right) }^2}
        \right) \\
T_2^{C_7^2} &=& {\frac{2}{3\hat{s}x}} \left( 1 + {{\hat{m}_s}^2}
\right) \left( \hat{\rho}_1 
      + 3\hat{\rho}_2 \right) \nonumber \\ 
&&+ {\frac{4}{3\hat{s}
     {x^2}}} \left( 1 + {{\hat{m}_s}^2} \right) 
     \left( 4\hat{\rho}_1 - \left( \hat{\rho}_1 + 3\hat{\rho}_2
       \right) v\cdot \hat{q} 
     \right) \nonumber \\ 
&&+ {\frac{8}{3
     \hat{s}{x^3}}} \left( 1 + {{\hat{m}_s}^2} \right) 
     \left( 3\hat{s}\hat{\rho}_2 + 2\left( 2\hat{\rho}_1 +
         3\hat{\rho}_2 \right) v\cdot \hat{q} -  
       2\left( 2\hat{\rho}_1 + 3\hat{\rho}_2 \right) {{\left( v\cdot \hat{q}
           \right) }^2} \right) \nonumber \\ 
&&- {\frac{32}{3\hat{s}{x^4}}} \left( 1 + {{\hat{m}_s}^2} \right)
\hat{\rho}_1\left( 1 - v\cdot \hat{q} \right) 
     \left( \hat{s} - {{\left( v\cdot \hat{q} \right) }^2} \right) \\
T_3^{C_7^2} &=& -{\frac{2}{3\,
     {{\hat{s}}^2}\,x}} \left( 1 - {{\hat{m}_s}^2} \right)
       \left( \hat{\rho}_1 + 3\hat{\rho}_2 \right) 
     v\cdot \hat{q} \nonumber \\ 
&&- {\frac{2}{3\,
     {{\hat{s}}^2}{\,x^2}}}\left( 1 -
           {{\hat{m}_s}^2} \right)
       v\cdot \hat{q}
     \left( \left( 8 - \hat{s} \right) 
        \hat{\rho}_1 - 3\hat{s}\hat{\rho}_2 - 
       2\left( \hat{\rho}_1 + 3\hat{\rho}_2 \right) 
        v\cdot \hat{q} \right)  \nonumber \\ 
&&+ {\frac{8}{3\,{{\hat{s}}^2}\,
     {x^3}}}\left( 1 - {{\hat{m}_s}^2} \right)
       \left( 1 - v\cdot \hat{q} \right) \nonumber\\
&&\;\;\;\;\;\;\;\;\;\;\;\;\;\;\;\;\;
     \left( \hat{s}
        \left( 2\hat{\rho}_1 + 3\hat{\rho}_2 \right) 
        + \hat{s}\left( \hat{\rho}_1 + 
          3\hat{\rho}_2 \right) v\cdot \hat{q} - 
       2\left( 2\hat{\rho}_1 + 3\hat{\rho}_2 \right)
          {{\left( v\cdot \hat{q} \right) }^2}
        \right)  \nonumber \\ 
&&- {\frac{16 }{3\,{{\hat{s}}^2}\,
     {x^4}}}\left( 1 - {{\hat{m}_s}^2} \right)
       \left( 1 - 
       v\cdot \hat{q} \right) \hat{\rho}_1
     \left( {{\hat{s}}^2} - 
       2\hat{s}\,v\cdot \hat{q} - 
       \hat{s}{{\left( v\cdot \hat{q} \right)
             }^2} + 
       2{{\left( v\cdot \hat{q} \right) }^3}
        \right)\\
T_1^{C_7 (C_9\pm C_{10})} &=& -{\frac{1}{x}} \left(\hat{\rho}_1 + 3\hat{\rho}_2\right) \nonumber \\ 
&&- {\frac{1}
     {3\,\hat{s}\,{x^2}}} \left( 2 \left( 3 + {{\hat{m}_s}^2} \right)
        \hat{s}\left( \hat{\rho}_1 + 
        3\hat{\rho}_2 \right)  - 
     6\left( 4\hat{\rho}_2 + 
        \hat{s}\left( \hat{\rho}_1 + 
           3\hat{\rho}_2 \right)  \right) 
      v\cdot \hat{q} \right. \nonumber\\
&&\left.\;\;\;\;\;\;\;\;\;\;\;
+ 
     2\left( 3 - {{\hat{m}_s}^2} \right) 
      \left( \hat{\rho}_1 + 3\hat{\rho}_2 \right) 
      {{\left( v\cdot \hat{q} \right) }^2} \right)\nonumber \\ 
&&+ {\frac{8}{3\,\hat{s}\,{x^3}}} \left( \left( 1 - 
          {{\hat{m}_s}^2} \right) \hat{\rho}_1 - 
       3{{\hat{m}_s}^2}\hat{\rho}_2 \right) 
     \left( 1 - v\cdot \hat{q} \right)  
     \left( \hat{s} -  
       {{\left( v\cdot \hat{q} \right) }^2} 
        \right)  \nonumber \\  
&&- {\frac{16}{3\,\hat{s}\,{x^4}} 
   } \hat{\rho}_1 
     \left( 1 - v\cdot \hat{q} \right)  
     \left( {{\hat{s}}^2} -  
       \left( 1 - {{\hat{m}_s}^2} \right)  
        \hat{s}\,v\cdot \hat{q} -  
       \hat{s}{{\left( v\cdot \hat{q} \right) 
             }^2} \right.\nonumber\\
&&\left.\;\;\;\;\;\;\;\;\;\;\;\;\;\;\;\;\;\;\;
+
       \left( 1 - {{\hat{m}_s}^2} \right)  
        {{\left( v\cdot \hat{q} \right) }^3} 
        \right) \\ 
T_2^{C_7 (C_9\pm C_{10})} &=& {\frac{4}{3\,{x^2}}} \left( \hat{\rho}_1 +
  3\hat{\rho}_2 \right) + 
{\frac{8}{\,{x^3}}} \hat{\rho}_2\left( {{\hat{m}_s}^2} + v\cdot
  \hat{q} \right) \\ 
T_3^{C_7 (C_9\pm C_{10})} &=& {\frac{2}{3\, 
     \hat{s}\,{x^2}}}\left( 12\hat{\rho}_2 -  
       \left( 3 + {{\hat{m}_s}^2} \right)  
        \left( \hat{\rho}_1 + 3\hat{\rho}_2 \right)  
        v\cdot \hat{q} \right) \nonumber\\
&&  - {\frac{8}{3\,\hat{s}\, 
     {x^3}}}\left( \hat{\rho}_1 +  
       {{\hat{m}_s}^2}\hat{\rho}_1 +  
       3{{\hat{m}_s}^2}\hat{\rho}_2 \right)  
     \left( 1 - v\cdot \hat{q} \right)  
     v\cdot \hat{q} \nonumber \\ 
&&+ {\frac{16}{3\,\hat{s}\,{x^4}} \left( 1 + {{\hat{m}_s}^2}
     \right)  
       \hat{\rho}_1\left( 1 -  
       v\cdot \hat{q} \right)  
     \left( \hat{s} -  
       {{\left( v\cdot \hat{q} \right) }^2} 
        \right) 
   } 
\end{eqnarray} 

\section{The dilepton invariant mass spectrum with full mass dependence}
\label{specwithmass}

In this Appendix we present the dilepton invariant mass spectrum for a
finite $s$-quark mass $m_s$. The spectrum originating from
operators of 
dimension $d \le 5$ has been presented in Eq. (47) of
\cite{ahhm_btosee_lepton}. The 
contributions from the time-ordered operators ${\cal T}_{1-4}$ can be 
obtained by making the replacement (\ref{tsubs}) in this
equation. Since the dilepton 
invariant mass distribution is independent of the definition of the
four velocity of the heavy quark, the contribution
proportional to $\rho_2$ is related by reparameterization invariance
\cite{repara} to the $\lambda_2$ contribution. It can be obtained by
the replacement
$$\lambda_2 \rightarrow \lambda_2 - \frac{\rho_2}{m_b},$$
in the results of \cite{ahhm_btosee_lepton}.

Thus, the only term we have to add to the existing literature to
obtain the complete expression including all $1/m_b^3$ contributions
is the term originating from the Darwin operator whose matrix element
is $\rho_1$. This contribution is given by
\begin{eqnarray}
\frac{d{\cal B}_{\rho_1}}{d\hat{s}} &=& {\cal B}_0 \rho_1 \Bigg[ \bigg( I^{(C_9^2 + C_{10}^2)}\,
\left(|C_9^{\rm eff}|^2 + C_{10}^2\right) + I^{C_7^2} \,|C_7^{\rm
eff}|^2  + I^{C_7 C_9} \,C_7^{\rm eff}
Re\left(C_9^{\rm eff}\right) \bigg) \left[ \frac{1}{\hat{u}^3(\hat{s},
\hat{m}_s)} \right]_*\, \nonumber  \\
   &+&  I_\delta \, \delta(\hat{s}_u - \hat{s}) \Bigg]\,,
\end{eqnarray}
with limits of integration defined in (\ref{lepton_phase})
\begin{equation}
\hat{s}_l \le \hat{s} \le \hat{s}_u\,,
\end{equation}
with
\begin{equation}
\hat{s}_l = 4 \hat{m}_l^2, \qquad \hat{s}_u = (1-\hat{m}_s)^2\,.
\end{equation}

The function $\hat{u}(\hat{s}, \hat{m}_s) = \sqrt{(\hat{s} - (1-\hat{m}_s)^2)
(\hat{s} - (1+\hat{m}_s)^2)}$ is singular at the upper limit of integration.
To regulate this divergence we defined a ``star function''
\begin{equation}
\left[F(\hat{s}, \hat{m}_s)\right]_* = \lim_{\beta\to 0}
\left\{F(\hat{s}, \hat{m}_s)  \theta(\hat{s}_u - \hat{s} - \beta) -
\delta( \hat{s}_u- \hat{s} - \beta) \int_{\hat{s}_l}^{\hat{s}_u -
\beta} \!\!\!\!\!\!\!d \hat{s} \,F(\hat{s}, \hat{m}_s) \right\}.
\end{equation}
This ``star function'' is analogous to the common plus distribution.

The functions appearing above are
\begin{eqnarray}
I^{(C_9^2 + C_{10}^2)} &=& - \frac{2}{9} \Bigg[
{\left( 1 - \hat{s} \right) }^2\,
   \left( 11 + 13\,\hat{s} - 9\,{\hat{s}}^2 + 
     23\,{\hat{s}}^3 + 10\,{\hat{s}}^4 \right)  \nonumber\\
&& - 
  \left( 50 + 37\,\hat{s} + 48\,{\hat{s}}^2 + 
     38\,{\hat{s}}^3 + 70\,{\hat{s}}^4 + 
     45\,{\hat{s}}^5 \right) \,{\hat{m}_s}^2 \nonumber\\
&& + 
  \left( 85 + 150\,\hat{s} + 216\,{\hat{s}}^2 + 
     178\,{\hat{s}}^3 + 75\,{\hat{s}}^4 \right) \,
   {\hat{m}_s}^4 - 2\,\left( 30 + 69\,\hat{s} + 
     72\,{\hat{s}}^2 + 25\,{\hat{s}}^3 \right) \,
   {\hat{m}_s}^6 \nonumber\\
&& + \left( 5 + 19\,\hat{s} \right) \,
   {\hat{m}_s}^8 + \left( 14 + 15\,\hat{s} \right) \,
   {\hat{m}_s}^{10} - 5\,{\hat{m}_s}^{12}\Bigg]\\
I^{C_7^2} &=& - \frac{8}{9 \hat{s}} \Bigg[
{\left( 1 - \hat{s} \right) }^2\,
   \left( 22 - 7\,\hat{s} + 9\,{\hat{s}}^2 + 
     19\,{\hat{s}}^3 + 5\,{\hat{s}}^4 \right)  \nonumber\\
&& - 
  \left( 78 - 34\,\hat{s} - 105\,{\hat{s}}^2 + 
     376\,{\hat{s}}^3 - 40\,{\hat{s}}^4 + 
     18\,{\hat{s}}^5 - 5\,{\hat{s}}^6 \right) \,
   {\hat{m}_s}^2 \nonumber\\
&& + \left( 70 + 67\,\hat{s} - 
     222\,{\hat{s}}^2 - 188\,{\hat{s}}^3 - 
     32\,{\hat{s}}^4 - 15\,{\hat{s}}^5 \right) \,
   {\hat{m}_s}^4 \nonumber\\
&& + \left( 50 + 60\,\hat{s} + 
     258\,{\hat{s}}^2 + 136\,{\hat{s}}^3 \right) \,
   {\hat{m}_s}^6 - \left( 110 + 157\,\hat{s} + 
     111\,{\hat{s}}^2 - 50\,{\hat{s}}^3 \right) \,
   {\hat{m}_s}^8 \nonumber\\
&& + \left( 38 + 2\,\hat{s} - 
     75\,{\hat{s}}^2 \right) \,{\hat{m}_s}^{10} + 
  9\,\left( 2 + 5\,\hat{s} \right) \,{\hat{m}_s}^{12} - 
  10\,{\hat{m}_s}^{14}\Bigg]\\
I^{C_7 C_9} &=& - \frac{8}{3} \Bigg[
{\left( 1 - \hat{s} \right) }^2\,
   (3 - \hat{s} + 17\,{\hat{s}}^2 - 3\,{\hat{s}}^3) \nonumber\\
&& - 
  \left( 10 - 13\,\hat{s} + 56\,{\hat{s}}^2 + 
     58\,{\hat{s}}^3 - 10\,{\hat{s}}^4 - 5\,{\hat{s}}^5
     \right) \,{\hat{m}_s}^2 \nonumber\\
&& + 
  \left( 5 - 22\,\hat{s} + 12\,{\hat{s}}^2 - 
     34\,{\hat{s}}^3 - 25\,{\hat{s}}^4 \right) \,
   {\hat{m}_s}^4 + 2\,\left( 10 + 29\,\hat{s} + 
     36\,{\hat{s}}^2 + 25\,{\hat{s}}^3 \right) \,
   {\hat{m}_s}^6 \nonumber\\
&& - \left( 35 + 67\,\hat{s} + 
     50\,{\hat{s}}^2 \right) \,{\hat{m}_s}^8 + 
  \left( 22 + 25\,\hat{s} \right) \,{\hat{m}_s}^{10} - 
  5\,{\hat{m}_s}^{12}\Bigg]\\
I_\delta &=& - \frac{16(1-\hat{m}_s^2)^5}{3 \sqrt{1-4\hat{m}_s^2}}\left(C_{10}^2+
  \left(C_9^{\mbox{eff}} (\hat{s})+2
  C_7^{\mbox{eff}}\right)^2 
\right) 
\end{eqnarray}
Notice that these terms correctly reproduce the expression (\ref{eqn:dbds0})
in the limit $\hat{m}_s \to 0$.

\section{The moments up to ${\cal O}(1/m_b^3)$ with a cut on the 
  dilepton invariant mass}\label{parton_moments} 

We write the moments in the form 
\begin{equation} 
M^{(m,n)} = \frac{{\cal B}_0}{{\cal B}_\chi} \left(C_{10}^2
  M^{(m,n)}_{10,10} + 
|C_7^{\mbox{eff}}|^2 M^{(m,n)}_{7,7} + M^{(m,n)}_{9,9} + C_7^{\mbox{eff}} 
M^{(m,n)}_{7,9} \right), 
\end{equation}  
where ${\cal B}_\chi$ is given in Eq.~(\ref{bchi}).  
The coefficient $C_9^{\mbox{eff}}$ depends on the parameters $x_0$ 
and $s_0$ as explained in section \ref{section_formalism}, so we express 
the moments $M^{(m,n)}_{9,9}$ and $M^{(m,n)}_{7,9}$ as integrals which
we evaluate numerically,    
\begin{eqnarray} 
M^{(m,n)}_{9,9} &=&   
\frac{16}{2M_B} 
\left(-\frac{1}{\pi} \right) \int_{\hat{m}_s}^{\frac{1}{2}(1-\chi)} \, dx_0
\int_{\hat{m}_s^2}^{x_0^2} \, d\hat{s}_0 \, x_0^m\, \hat{s}_0^n
\Bigg[ \sqrt{x_0^2 - \hat{s}_0} \;\mbox{Im}  
\Bigg\{ \Bigg[ 2 (1 - 2 x_0 + \hat{s}_0) 
    T_1^{{(C_9\pm C_{10})^2}}
   \nonumber \\  
&\qquad& +  \frac{x_0^2 - 
      \hat{s}_0}{3} T_2^{{(C_9\pm C_{10})^2}}
  \Bigg] \Bigg\} |C_9^{\mbox{eff}}(x_0,\hat{s}_0)|^2 \Bigg] \nonumber\\ 
M^{(m,n)}_{7,9} &=&  
\frac{16}{2M_B} 
\left(-\frac{1}{\pi} \right) \int_{\hat{m}_s}^{\frac{1}{2}(1-\chi)} \, dx_0
\int_{\hat{m}_s^2}^{x_0^2} \, d\hat{s}_0 \, x_0^m\, \hat{s}_0^n
\Bigg[ \sqrt{x_0^2 - \hat{s}_0} \;\mbox{Im}  
\Bigg\{ \Bigg[ 2 (1 - 2 x_0 + \hat{s}_0) 
    T_1^{C_7 (C_9\pm C_{10})}  
   \nonumber \\  
&& \qquad + \frac{x_0^2 - 
      \hat{s}_0}{3} T_2^{C_7 (C_9\pm C_{10})}
  \Bigg] \Bigg\}Re\left[ C_9^{\mbox{eff}}(x_0,\hat{s}_0)\right] \Bigg] 
\end{eqnarray}   
For the other contributions we find 
\begin{eqnarray} 
M^{(1,0)}_{10,10} &=&  \frac{{\left( 1 - \chi \right) }^4\,
    \left( 7 + 8\,\chi \right) }{30}  + \frac{\lambda_1}{m_b^2} \left(
    \frac{{\left( 1 - \chi \right) }^3\,
      \left( 1 + \chi \right) }{3}  \right) \nonumber\\
&&\quad
  - \frac{\lambda_2}{m_b^2}
  \left(  \frac{2\,{\left( 1 - \chi \right) }^2\,\chi\,
    \left( 1 + 6\,\chi - 4\,{\chi}^2 \right) }{3}  \right)
\nonumber\\ 
&&\quad
  - \frac{\rho_1}{m_b^3} \left( \frac{67 - 30\,\chi + 30\,{\chi}^3 -
      35\,{\chi}^4 -  
    32\,{\chi}^5}{45}   \right) \nonumber\\ 
&&\quad
+ 
  \frac{\rho_2}{m_b^3} \left(  \frac{1 + 10 \,\chi - 50\,{\chi}^3 +
    55\,{\chi}^4 - 16\,\chi^5}{5} \right) \\
M^{(1,0)}_{7,7} &=&  -\frac{2\,\left( 41 - 60\,\chi +
      18\,{\chi}^2 + 4\,{\chi}^3 -  
      3\,{\chi}^4 + 24\,\log (\chi) \right) }{9}  \nonumber\\
&&\quad
-
  \frac{\lambda_1}{m_b^2} \left( \frac{8\,\left( 8 - 9\,\chi +
        {\chi}^3 + 6\,\log (\chi) \right) 
      }{9}  \right) \nonumber\\
&&\quad
  + \frac{\lambda_2}{m_b^2}
  \left(  \frac{4\,\left( 7 - 2\,\chi - 10\,{\chi}^3 + 5\,{\chi}^4 + 
      12\,\log (\chi) \right) }{3}  \right) \nonumber\\
&&\quad
  + \frac{\rho_1}{m_b^3} \left(  \frac{8\,\left( 2 - 27\,\chi +
        19\,{\chi}^3 + 6\,{\chi}^4 +  
      18\,\log (\chi) \right) }{27}  \right) \nonumber\\ 
&&\quad
+ 
  \frac{\rho_2}{m_b^3} \left(  \frac{8\,\left( 13 - 27\,\chi +
        23\,{\chi}^3 - 9\,{\chi}^4 -  
      6\,\log (\chi) \right) }{9}   \right) \\
M^{(2,0)}_{10,10} &=&  \frac{{\left( 1 - \chi \right) }^5\,
      \left( 4 + 5\,\chi \right)  }{45}  +
  \frac{\lambda_1}{m_b^2} \left(
\frac{{\left( 1 - \chi \right) }^4\,
    \left( 43 + 67\,\chi + 25\,{\chi}^2 \right) }{270}  \right)
\nonumber\\ 
&&\quad
  + \frac{\lambda_2}{m_b^2}
  \left(  \frac{{\left( 1 - \chi \right) }^3\,
      \left( 13 + 24\,\chi - 222\,{\chi}^2 + 125\,{\chi}^3
        \right) }{90}  \right) \nonumber\\
&&\quad
  - \frac{\rho_1}{m_b^3} \left( \frac{{\left( 1 - \chi \right) }^2\,
      \left( 24 - 27\,\chi - 3\,{\chi}^2 + 191\,{\chi}^3 + 
        175\,{\chi}^4 \right)  }{270}   \right) \nonumber\\ 
&&\quad
- 
  \frac{\rho_2}{m_b^3} \left(  \frac{{\left( 1 - \chi \right) }^3\,
    \left( 14 - 63\,\chi - 306\,{\chi}^2 + 175\,{\chi}^3 \right)
      }{90}   \right) \\
M^{(2,0)}_{7,7} &=&  -\frac{2\,\left( 119 - 210\,\chi +  
      120\,{\chi}^2 - 20\,{\chi}^3 - 15 {\chi}^4 + 
      6\,{\chi}^5 + 60\,\log (\chi) \right) }{45}
  \nonumber\\
&&\quad\left.
- \frac{\lambda_1}{m_b^2} \left(  \frac{127 - 150\,\chi -
      12\,{\chi}^2 + 44\,{\chi}^3 -  
    3\,{\chi}^4 - 6\,{\chi}^5 + 84\,\log (\chi)}{27} \right)
\right.\nonumber\\ 
&&\quad\left. 
  + \frac{\lambda_2}{m_b^2}
  \left( \frac{9 - 2\,\chi + 4\,{\chi}^2 - 36\,{\chi}^3 + 35\,{\chi}^4 - 
    10\,{\chi}^5 + 12\,\log (\chi)}{3}   \right) \right. \nonumber\\
&&\quad\left.
  + \frac{\rho_1}{m_b^3} \left( \frac{127 - 210\,\chi - 24\,{\chi}^2 +
      188\,{\chi}^3 -  
    39\,{\chi}^4 - 42\,{\chi}^5 + 60\,\log (\chi)}{27}   \right)
\right.\nonumber\\  
&&\quad
+ 
  \frac{\rho_2}{m_b^3} \left(  \frac{ 39 - 66\,\chi - 8\,\chi^2 +
    84\,{\chi}^3 -  
  63\,{\chi}^4 + 14\,\chi^5 + 12\,\log (\chi)}{3}   \right)
\\ 
M^{(0,1)}_{10,10} &=&  \frac{\lambda_1}{m_b^2} \left(  
\frac{{\left( 1 - \chi \right) }^3\,
      \left( 13 + 19\,\chi + 8\,{\chi}^2 \right)  }{30}
\right)\nonumber\\ 
&&\quad\left.
  + \frac{\lambda_2}{m_b^2}
  \left(  \frac{{\left( 1 - \chi \right) }^3\,
    \left( 3 + 13\,\chi - 8\,{\chi}^2 \right) }{6}  \right)
\right. \nonumber\\ 
&&\quad\left.
  + \frac{\rho_1}{m_b^3} \left(  \frac{{\left( 1 - \chi \right) }^2\,
    \left( 177 + 254\,\chi + 201\,{\chi}^2 + 88\,{\chi}^3
      \right) }{90}  \right) \right.\nonumber\\ 
&&\quad
+ 
  \frac{\rho_2}{m_b^3} \left(  \frac{{\left( 1 - \chi \right)
          }^2\, 
      \left( 1 - 18\,\chi - 47\,{\chi}^2 + 24\,{\chi}^3 \right) 
      }{10}   \right) \\
M^{(0,1)}_{7,7} &=&  - \frac{\lambda_1}{m_b^2} \left(
    \frac{2\,\left( 23 - 12\,\chi - 18\,{\chi}^2 + 4\,{\chi}^3 +  
      3\,{\chi}^4 + 24\,\log (\chi) \right) }{9}   \right)
\nonumber\\ 
&&\quad\left.
  - \frac{\lambda_2}{m_b^2}
  \left(  \frac{2\,\left( 31 - 28\,\chi - 18\,{\chi}^2 + 20\,{\chi}^3 - 
      5\,{\chi}^4 + 24\,\log (\chi) \right) }{3}  \right)
\right. \nonumber\\ 
&&\quad\left.
  + \frac{\rho_1}{m_b^3} \left( \frac{2\,\left( 77 - 132\,\chi -
        54\,{\chi}^2 + 76\,{\chi}^3 +  
      33\,{\chi}^4 - 120\,\log (\chi) \right) }{27}   \right)
\right.\nonumber\\  
&&\quad
+ 
  \frac{\rho_2}{m_b^3} \left(  \frac{2\,\left( 281 - 324\,\chi -
        54\,{\chi}^2 +  
      124\,{\chi}^3 - 27\,{\chi}^4 + 168\,\log (\chi) \right) }
    {9}   \right) \\
M^{(0,2)}_{10,10} &=&  -\frac{\lambda_1}{m_b^2} \left(
    \frac{4\,{\left( 1 - \chi \right) }^5\, 
    \left( 4 + 5\,\chi \right) }{135}  \right) \nonumber\\
&&\quad\left.
  - \frac{\rho_1}{m_b^3} \left(  \frac{2\,{\left( 1 - \chi \right)
        }^4\, 
    \left( 31 + 64\,\chi + 40\,{\chi}^2 \right) }{135}  \right)
\right.\nonumber\\  
&&\quad
+ 
  \frac{\rho_2}{m_b^3} \left(  \frac{2\,{\left( 1 - \chi \right) }^4\,
    \left( 1 - 26\,\chi + 10\,{\chi}^2 \right) }{45}   \right)
\\ 
M^{(0,2)}_{7,7} &=&  \frac{\lambda_1}{m_b^2} \left(
    \frac{8\,\left( 119 - 210\,\chi + 120\,{\chi}^2 -  
      20\,{\chi}^3 - 15\,{\chi}^4 + 6\,{\chi}^5 + 
      60\,\log (\chi) \right) }{135}  \right) \nonumber\\
&&\quad\left.
  + \frac{\rho_1}{m_b^3} \left(  \frac{8\,\left( 139 - 60\,\chi -
        210\,{\chi}^2 +  
      140\,{\chi}^3 + 15\,{\chi}^4 - 24\,{\chi}^5 + 
      120\,\log (\chi) \right) }{135}  \right) \right.\nonumber\\ 
&&\quad
- 
  \frac{\rho_2}{m_b^3} \left(  \frac{16\,\left( 73 - 165\,\chi +
        165\,{\chi}^2 -  
      100\,{\chi}^3 + 30\,{\chi}^4 - 3\,{\chi}^5 + 
      30\,\log (\chi) \right) }{45}   \right) \\
M^{(1,1)}_{10,10} &=&  \frac{\lambda_1}{m_b^2} \left(  
\frac{{\left( 1 - \chi \right) }^4\,
    \left( 23 + 62\,\chi + 50\,{\chi}^2 \right) }{270} \right)
\nonumber\\ 
&&\quad\left.
  + \frac{\lambda_2}{m_b^2}
  \left(  \frac{{\left( 1 - \chi \right) }^4\,
      \left( 13 + 82\,\chi - 50\,{\chi}^2 \right)  }{
    90}  \right) \right. \nonumber\\
&&\quad\left.
  + \frac{\rho_1}{m_b^3} \left(  \frac{{\left( 1 - \chi
          \right) }^3\, 
      \left( 71 + 183\,\chi + 276\,{\chi}^2 + 190\,{\chi}^3
        \right)  }{270}  \right) \right.\nonumber\\ 
&&\quad
- 
  \frac{\rho_2}{m_b^3} \left(  \frac{{\left( 1 - \chi \right) }^3\,
    \left( 13 + 9\,\chi - 252\,{\chi}^2 + 110\,{\chi}^3 \right) 
    }{90}   \right) \\
M^{(1,1)}_{7,7} &=&  -\frac{\lambda_1}{m_b^2} \left( 
\frac{4\,\left( 2 + 15\,\chi - 33\,{\chi}^2 + 16\,{\chi}^3 + 
      3\,{\chi}^4 - 3\,{\chi}^5 + 6\,\log (\chi) \right) }{27}
\right) \nonumber\\ 
&&\quad\left.
  - \frac{\lambda_2}{m_b^2}
  \left(  \frac{4\,\left( 10 - 13\,\chi - {\chi}^2 + 8\,{\chi}^3 - 
      5\,{\chi}^4 + {\chi}^5 + 6\,\log (\chi) \right) }{3}  \right)
\right. \nonumber\\ 
&&\quad\left.
  + \frac{\rho_1}{m_b^3} \left( \frac{4\,\left( 62 - 75\,\chi -
        195\,{\chi}^2 +  
      280\,{\chi}^3 - 15\,{\chi}^4 - 57\,{\chi}^5 - 
      30\,\log (\chi) \right) }{135}   \right) \right.\nonumber\\ 
&&\quad
+ 
  \frac{\rho_2}{m_b^3} \left(  \frac{4\,\left( 257 - 285\,\chi -
        195\,{\chi}^2 +  
      400\,{\chi}^3 - 210\,{\chi}^4 + 33\,{\chi}^5 + 
      150\,\log (\chi) \right) }{45}   \right) 
\end{eqnarray}


\begin{thebibliography}{99}
\bibitem{grinstein_savage_wise}B.Grinstein, M.Savage, and M.B.Wise, \np{319}{1989}{271}.
\bibitem{misiak}M. Misiak, \np{393}{1993}{23}; {\bf Erratum}, 
   \np{439}{1995}{461}. 
\bibitem{buras-munz} A.J. Buras and M.M\"{u}nz, \prd{52}{1995}{186}.
\bibitem{newphysics} A.Ali, G.F.Giudice, and T.Mannel, Z.Phys.{\bf C67} (1995) 417; C.Huang, W.Liao, and Q.Yan, \prd{59}{1999}{011701}; P.Cho, M.Misiak, and D.Wyler, \prd{54}{1996}{3329}; J.Hewett and J.D.Wells, \prd{55}{1997}{5549}; T.Goto, Y.Okada, Y.Shimizu, and M.Tanaka, \prd{55}{1997}{4273}; E.Lunghi, A.Masiero, I.Scimemi, and L.Silvestrini, hep-ph/9906286. 
\bibitem{CLEOex}R. Ammar {\it et al.} (CLEO Collaboration),\prl{71}{1993}{674}.
\bibitem{CLEOin}M.S. Alam {\it et al.} (CLEO Collaboration), Phys. Rev. Lett. {\bf 74} (1995) 2885; hep-ph/9908022.
\bibitem{constraints} J.Hewett, hep-ph/9406302; T.G.Rizzo, \prd{58}{1998}{114014}; F.Borzumati and C.Greub, \prd{58}{1998}{074004}; W.de Boer {\it et al.}, \pl{438}{1998}{281}; H.Baer {\it et al.}, \prd{58}{1998}{015007}; M.Ciuchini {\it et al.}, \np{527}{1998}{21}; J.Agrawal {\it et al.}, \IJMP \blankref{A11} (1996) 2263; M.A.Diaz {\it et al.}, \np{551}{1999}{78}; G.V.Kranrotis, Z.Phys.{\bf C71} (1996) 163; A.L.Kagan and M.Neubert, {\it Eur.Phys.J.}{\bf C7} (1999) 5. 
\bibitem{experimental_search} S. Glenn {\it et al.} (CLEO Collaboration), \prl{80}{1998}{2289}; B.Abbott {\it et al.} (D{\O} Collaboration), \pl{423}{1998}{419}; C.Albajar {\it et al.} (UA1 Collaboration), \pl{262}{1991}{163}.  Also, a recent search in exclusive channels: T.Affolder {\it et al.} (CDF Collaboration), hep-ex/9905004.
\bibitem{ahhm_btosee_lepton}A. Ali, L.T.Handoko, G.Hiller, T.Morozumi, 
        \prd{55}{1997}{4105}.
\bibitem{ah_btosee_hadron} A. Ali and G. Hiller, \prd{58}{1998}{074001}. 
\bibitem{b->see}B.Grinstein, M.Savage, and M.B.Wise, \np{319}{1989}{271}.
\bibitem{lam1values} Z.Ligeti, Y.Nir, \prd{49}{1994}{4331}; M.Gremm, 
  A.Kapustin, Z.Ligeti, M.Wise, \prl{77}{1996}{20}; M.Neubert,
  \pl{389}{1996}{727}.
\bibitem{bauerburrell} C. Bauer and C.Burrell, \pl{469}{1999}{248}.
\bibitem{chay}J. Chay, H. Georgi and B. Grinstein, Phys. Lett. {\bf
  B247} (1990) 399.
\bibitem{inclusive}I.I. Bigi, M. Shifman, N.G. Uraltsev,
A. Vainshtein, Phys. Rev. Lett. {\bf 71}
(1993) 496, B. Blok, L. Koyrakh, M. Shifman,  A. Vainshtein,
Phys. Rev. {\bf D49} (1994) 3356, Erratum-ibid. {\bf D50} (1994) 3572,
A. V. Manohar, M. B. Wise, Phys. Rev. {\bf D49} (1994) 1310; I.I.Bigi, N.G.Uraltsev, A.I.Vainshtein, \pl{293}{1992}{430}; {\bf Erratum}, \pl{297}{1993}{477}.
\bibitem{fls1994} A.Falk, M.Luke, and M.Savage, \prd{49}{1994}{3367}. 
\bibitem{christian}C. Bauer, Phys. Rev. {\bf D57} (1998) 5611; {\bf
    Erratum} \prd{60}{1999}{099907}.
\bibitem{gremm}M. Gremm and A. Kapustin, Phys. Rev. {\bf D55} (1997) 6924.
\bibitem{paramd2}A.F. Falk and M.Neubert, Phys. Rev. {\bf D47} (1993) 2965;
  \prd{47}{1993}{2982}.
\bibitem{mannel}T. Mannel, Phys. Rev. {\bf D50} (1994) 428.
\bibitem{hadmass}A. F. Falk, M. Luke, M. Savage, Phys. Rev. {\bf D53} (1996) 2491.
\bibitem{vac_sat}
M.A. Shifman and M.B. Voloshin, Sov. J. Nucl. Phys. 45 (1987) 292;
M.B. Voloshin, N.G. Uraltsev, V.A. Khoze and M.A. Shifman,
Sov.~J.~Nucl.~Phys. 46 (1987) 112.
\bibitem{blok}B.~Blok, R.D.~Dikeman and M.~Shifman, Phys. Rev. {\bf D51}, 
        6167 (1995).
\bibitem{sumlogs}C. Bauer, A. Falk and M. Luke, \prd{54}{1996}{2097}.
\bibitem{OPEbreakdown} G.Buchalla and G.Isidori, \np{525}{1998}{333}. 
\bibitem{shape_func}M. Neubert, Phys. Rev. {\bf D49} (1994) 3367; Phys. Rev. {\bf D49} (1994) 4623. 
\bibitem{leptoncut} A. Falk and M. Luke, \prd{57}{1998}{424}.
\bibitem{repara} M.Luke and A.Manohar, \pl{286}{1992}{348}. 

\end{thebibliography}
\end{document}